%% file: main.tex
\documentclass[sigconf, nonacm, screen]{acmart}
\usepackage{tikz}
\DeclareMathAlphabet{\mathcal}{OMS}{cmsy}{m}{n}
\usepackage{amsthm}
\usepackage{textgreek}
\usepackage[english]{babel}

\newtheorem{lemma}{Lemma}[section] 
\newtheorem{theorem}{Theorem}[section]

\usepackage{mathtools}
\DeclarePairedDelimiter\ceil{\lceil}{\rceil}
\DeclarePairedDelimiter\floor{\lfloor}{\rfloor}

\usepackage{amsmath}
\usepackage{array, xspace, xcolor, algorithm}

\usepackage[noend]{algpseudocode}

\usepackage{adjustbox}
\usepackage{subcaption}
\usepackage{xspace}
\usepackage{enumitem}

\newcommand{\cab}{Cabinet\xspace}

\begin{document}

\title{\cab: Dynamically Weighted Consensus Made Fast\\ (Extended Version)}

\author{Gengrui Zhang}
\affiliation{%
  \institution{Concordia University}
}
\email{gengrui.zhang@concordia.ca}

\author{Shiquan Zhang}
\affiliation{%
  \institution{University of Toronto}
}
\email{shiquan.zhang@mail.utoronto.ca}

\author{Michail Bachras}
\affiliation{%
  \institution{University of Toronto}
}
\email{michalis.bachras@mail.utoronto.ca}

\author{Yuqiu Zhang}
\affiliation{%
  \institution{University of Toronto}
}
\email{quincy.zhang@mail.utoronto.ca}

\author{Hans-Arno Jacobsen}
\affiliation{%
  \institution{University of Toronto}
}
\email{jacobsen@eecg.toronto.edu}

\begin{abstract}
Conventional consensus algorithms, such as Paxos and Raft, encounter inefficiencies when applied to large-scale distributed systems due to the requirement of waiting for replies from a majority of nodes. To address these challenges, we propose \cab, a novel consensus algorithm that introduces dynamically weighted consensus, allocating distinct weights to nodes based on any given failure thresholds. \cab dynamically adjusts nodes' weights according to their responsiveness, assigning higher weights to faster nodes. The dynamic weight assignment maintains an optimal system performance, especially in large-scale and heterogeneous systems where node responsiveness varies. We evaluate \cab against Raft with distributed MongoDB and PostgreSQL databases using YCSB and TPC-C workloads. The evaluation results show that \cab outperforms Raft in throughput and latency under increasing system scales, complex networks, and failures in both homogeneous and heterogeneous clusters, offering a promising high-performance consensus solution.
\end{abstract}

\maketitle

\pagestyle{plain}

\ifdefempty{\vldbavailabilityurl}{}{
\vspace{.3cm}
\begingroup\small\noindent\raggedright\textbf{About this manuscript:}\\
This paper is an extended version of Cabinet. The extension is presented in Section~\ref{sec:subsec:correctness}, including the discussion of \cab's correctness for replication and leader election.
\endgroup
}

\ifdefempty{\vldbavailabilityurl}{}{
\vspace{.3cm}
\begingroup\small\noindent\raggedright\textbf{Artifacts Availability:}\\
\cab is implemented in Go and is fully open-source. The source code and deployment artifacts have been made available at \url{https://github.com/gengruizhang/cabinet}. 
\vspace{.6cm}
\endgroup
}

\section{Introduction}
\label{sec:introduction}

\begin{figure}
    \centering
    \includegraphics[width=0.8\linewidth]{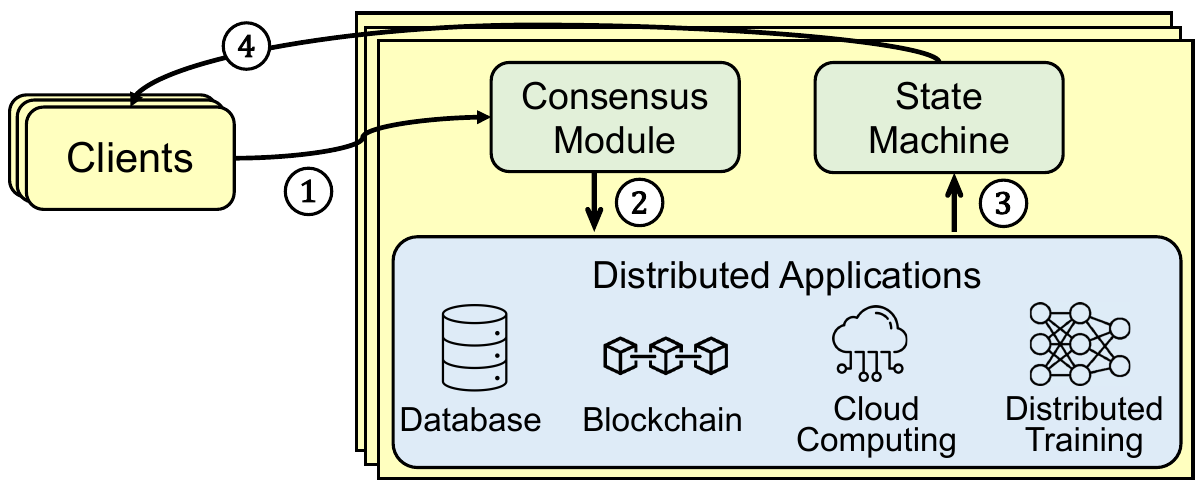}
    \caption{Structure of distributed consensus applications.}
    \label{fig:dappl}
\end{figure}

Consensus algorithms are crucial for ensuring data consistency across nodes in distributed applications. These applications, typically structured as depicted in Figure~\ref{fig:dappl}, consist of clients and services (coordinated nodes). Clients invoke a consensus task by sending requests to services. Then, the consensus algorithm coordinates nodes to agree on the order of the executions of these tasks. After that, nodes update their states and notify the client. With the rapid development of distributed applications, the scale of these systems has witnessed a significant upsurge~\cite{amiri2020modern}, including permissioned blockchains~\cite{androulaki2018hyperledger, diem}, distributed storage~\cite{ousterhout2015ramcloud, corbett2013spanner, apachecassandra}, cloud management~\cite{dockerswarm,etcd}, and distributed training systems~\cite{tensorflowdistributed, pytorchdistributed}. 

In particular, Paxos~\cite{leslie1998part}, Raft~\cite{ongaro2014search}, and their variants~\cite{howard2021fast, zhangcep, cassandrapaxos, zhang2019efficient} have been widely used to achieve quorum replication, in which consensus follows a \textit{simple-majority rule} and tolerates up to $f=\floor{\frac{n-1}{2}}$ benign failures in a system of $n$ nodes. However, as the quorum size grows linearly with the system size, simple-majority becomes inefficient when applied to large-scale systems, particularly those that are heterogeneous~\cite{zhao2018sdpaxos}, such as global-scale databases~\cite{180268, corbett2013spanner} and blockchains applications~\cite{androulaki2018hyperledger}.

The root cause lies in the fact that the majority rule requires collecting $f+1$ replies from a total of $n=2f+1$ nodes. As the system scales, the number of required replies increases linearly with the system size. For instance, in Google's Spanner's evaluation~\cite{180268}, despite the system scaling from tens to hundreds of nodes, the quorum size keeps growing (e.g., $51$ replies in $100$ nodes), leading to low performance, especially with high latency, though the probability of half of the nodes failing in the system is exceedingly low.

Furthermore, nodes in modern applications are often heterogeneous with varying configurations~\cite{barroso2019datacenter, mars2013whare}, with some nodes possessing stronger configurations, such as higher computational capabilities and faster memory access. When the stronger nodes constitute less than a majority, they are compelled to wait for the slower nodes, leading to system-wide performance degradation~\cite{sahoo2004failure, schroeder2009large, tiwari2015understanding}.

To tackle these challenges, we propose \cab, a novel consensus algorithm that introduces \textit{weighted consensus} for efficient and fast agreement. \cab offers a customizable failure threshold $t$, where $t$ can be any integer between $1 \leq t \leq \floor{\frac{n-1}{2}}$, along with a tailored weight scheme that assigns each node a distinct weight. \cab defines the $t+1$ nodes with the highest weights as the \textit{cabinet members} and enables \textit{weight quorums}, achieving system-wide agreement as soon as the cabinet members reach an agreement. 

In addition, the composition of cabinet members dynamically changes to maintain an optimal configuration. In each round, the leader reassigns weights to nodes based on their responsiveness. This ensures that faster nodes are assigned higher weights in the subsequent execution. The $t+1$ most responsive nodes in the current execution will become the cabinet members for the next execution.

\cab's weighted consensus achieves fast consensus by prioritizing faster nodes with higher weights. For example, in heterogeneous clusters, strong nodes can become cabinet members when they exhibit higher responsiveness, which prevents the system from being delayed by slower nodes. Furthermore, under unstable networks, the weight reassignment can promptly adjust weights by including the most responsive nodes to cabinet members and excluding slow nodes, thereby maintaining an optimal configuration. When node responsiveness fluctuates, the composition of cabinet members adapts accordingly to always include the $t+1$ most responsive nodes in the current round. 

Compared to conventional consensus, \cab's weighted consensus enables more flexible fault tolerance. It tolerates a minimum of $t$ nodes (i.e., the nodes of top $t$ weights fail) and a maximum of $n{-}t{-}1$ nodes (i.e., the nodes of top $t\!+\!1$ weights survive). In heterogeneous clusters, since weak nodes often suffer from limited resources, overutilization, and aging hardware, they are statistically more likely to fail than strong nodes~\cite{sahoo2004failure}. By assigning lower weights to weak nodes, \cab can effectively contribute to a more pragmatic and efficient fault tolerance strategy.

We implemented \cab based on Raft and compared their performance using the YCSB~\cite{cooper2010benchmarking} and TPC-C~\cite{tpcc} workloads on distributedly deployed MongoDB~\cite{mongodb} and PostgreSQL~\cite{postgresql}. Our evaluation encompassed factors such as heterogeneity, cluster scalability, complex network conditions, and node failures. \cab consistently delivered high performance, including higher throughput and lower latency, across all tested cases with remarkable efficiency, robustness, and adaptability, making it an ideal choice for large-scale, heterogeneous distributed systems.

To summarize, \cab makes the following contributions:

\begin{enumerate}

\item It defines the properties and fundamentals of weighted consensus and formulates the establishment and calculation of weight schemes that achieve fast agreement with any given failure threshold.

\item It allows a flexible fault tolerance and yields the option of choosing the failure threshold ($t$) to applications. It can tolerate a minimum of $t$ failures (when high-weight nodes fail) in the worst case and $n-t-1$ failures (when high-weight nodes survive) in the best case.

\item It enables fast agreement with weight quorums and dynamically rearranges node weights based on responsiveness, which empowers the system to maintain optimal performance under network changes and failures.

\item It proposes a general-purpose benchmark framework for evaluating distributed consensus applications. Compared to conventional consensus, \cab achieves higher throughput and lower latency when applied to database applications under both YCSB and TPC-C workloads.

\end{enumerate}

The rest of the paper is organized as follows: \S\ref{sec:background} presents the background and related work; \S\ref{sec:weightscheme} defines weighted consensus and weight schemes; \S\ref{sec:cabalgo} describes the \cab consensus algorithm; and \S\ref{sec:evaluation} reports the evaluation result.

\section{Background and Related Work}
\label{sec:background}
Consensus algorithms are closely associated with state machine replication (SMR) in distributed systems~\cite{schneider1990implementing}. They coordinate a group of nodes to agree on a value, even in the presence of failures. Consensus algorithms are majorly categorized into two classes: the one that tolerates \textit{benign failures} (e.g., Paxos~\cite{leslie1998part} and Raft~\cite{ongaro2014search}), such as crashes or network delays, and the other one that tolerates \textit{Byzantine failures}~\cite{zhang2024reaching} (e.g., PBFT~\cite{castro1999practical}, Hotstuff~\cite{yin2019hotstuff}, and PrestigeBFT~\cite{zhang2024prestigebft}), such as arbitrary behavior where nodes may behave maliciously~\cite{lamport1982byzantine, zhang2021prosecutor}. This paper focuses on the consensus algorithms that tolerate benign failures (i.e., non-Byzantine failures). Paxos and Raft serve as exemplary consensus algorithms and have been widely deployed in numerous applications. 

\textbf{Majority quorums and optimizations.} In the Paxos/Raft consensus protocol family~\cite{leslie1998part, marandi2010ring, lamport2009vertical, lamport2004cheap, shi2016cheap, ongaro2014search}, majority quorums, adhering to the simple majority rule~\cite{agrawal1992generalized}, are the most common quorum replication building blocks. The quorum size typically corresponds to a majority of the total number of nodes; i.e., $\floor{\frac{n}{2}}+1$ in a system of $n$ nodes~\cite{herlihy1986quorum}. Majority quorums use the concept of simple majority to prevent nodes from making conflicting decisions and achieve fault tolerance by tolerating $\floor{\frac{n-1}{2}}$ failures.

However, majority quorums become inefficient as distributed applications continue to grow in scale, such as distributed databases~\cite{amiri2020modern, netflix1, netflix2}, blockchains~\cite{androulaki2018hyperledger, shamis2022ia,russinovich2019ccf}, and distributed training~\cite{verbraeken2020survey, tensorflowdistributed, pytorchdistributed}.
Their consensus threshold (i.e., $\floor{\frac{n}{2}}+1$) grows linearly as the scale of the system increases. However, in modern large-scale computer systems with a significant number of nodes (e.g., $n>50$ nodes), the likelihood of half the nodes crashing simultaneously is highly improbable~\cite{sahoo2004failure, schroeder2009large, tiwari2015understanding}. While maintaining fault tolerance remains critical, it is essential to reassess the quorum size and strike a more appropriate balance between fault tolerance and efficiency~\cite{brewer2012cap}.

Tailoring to specific application scenarios, numerous consensus protocols have proposed optimizations, especially based on Paxos~\cite{ports2015designing, charapko2021pigpaxos, li2016just, bolosky2011paxos, barcelona2008mencius, moraru2012egalitarian}. For example, Vertical Paxos~\cite{lamport2009vertical} enables reconfiguration that moves “vertically” to change the leader node, while each distinct vertical configuration, managed by different proposers, conducts Paxos instances along a "horizontal" axis. Furthermore, Flexible Paxos~\cite{howard2017flexible, howard2022relaxed} introduces simple quorums and grid quorums that increase the size of propose quorums ($Q_1$) and reduce that of accept quorums ($Q_2$) where $|Q_1|+|Q_2|>N$ (the number of nodes). With grid quorums, it can reduce the average quorum size to $N_1+N_2/2$.

\begin{figure}
    \centering
    \includegraphics[width=0.95\linewidth]{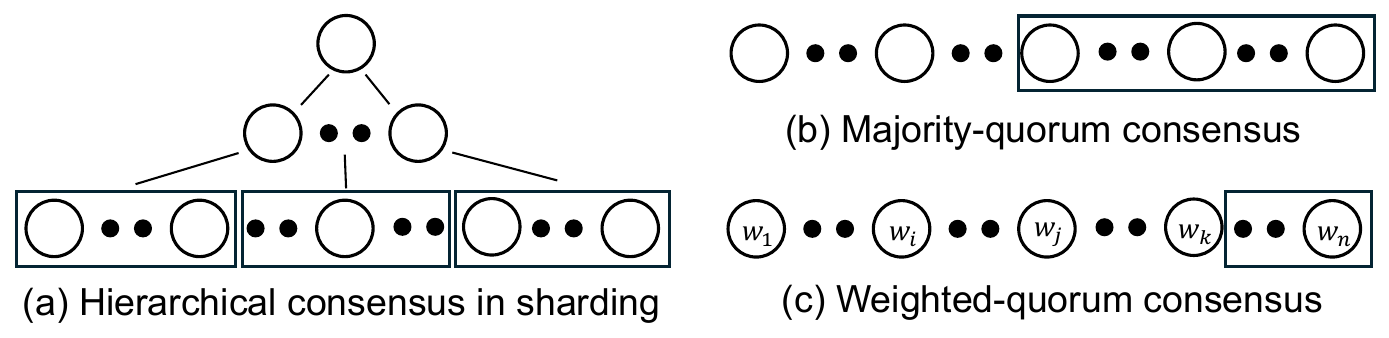}
    \caption{Illustrations of (a) hierarchical, (b) majority, and (c) weighted consensus mechanisms.}
    \label{fig:rw-comp}
\end{figure}

\textbf{Sharding on scaling systems.}
When scaling a system, sharding—partitioning a large group of nodes into hierarchical quorums—is often applied, as exemplified by Google Spanner~\cite{corbett2013spanner}. Compared to majority-quorum consensus, hierarchical quorum consensus (HQC) allows consensus decisions to be made simultaneously within groups. Once a consensus decision is reached at the child-level groups, parent groups aggregate these decisions up to the root group (shown in Figure~\ref{fig:rw-comp}). HQC reduces the quorum size needed for each decision-making process, achieving a quorum size of $n^{0.63}$\cite{kumar1991algorithm}, in contrast to the $\ceil{\frac{n+1}{2}}$ size required in majority quorums. HQC is also listed as an option for consensus mechanisms in Zookeeper\cite{zkweight}.

However, HQC introduces significant latency, as decisions must traverse multiple rounds with respect to the height of the hierarchical tree~\cite{corbett2013spanner, spannersupport}. In contrast, non-partitioned consensus algorithms, such as Paxos and Raft, typically require only two rounds of message-passing to reach a decision~\cite{lamport2001paxos, ongaro2014search}. However, they become inefficient in large-scale distributed replication due to its larger quorum size requirement. Thus, the inefficiency of large-scale replication remains a critical challenge in practice~\cite{Dynamoat15}.

\textbf{When nodes are not born equal.} In today's distributed applications, heterogeneity is a common feature among nodes~\cite{zhang2023efficient}. For instance, in distributed databases and blockchain applications, heterogeneous nodes often coexist with varying configurations among different paricipating entities. Moreover, even within a single entity, heterogeneity prevails due to diverse hardware acquisitions and varying VM setups. Additionally, nodes frequently exhibit divergent responsiveness because of their network conditions, resulting in varying response times, with some nodes being faster at one moment and slower at another. These features heavily impact the overall quality of major quorum consensus, as strong nodes are compelled to wait for weak nodes to reach consensus. Furthermore, weak nodes in heterogeneous clusters statistically experience more failures than strong nodes because of their limited resources, overutilization, and aging hardware~\cite{sahoo2004failure}. 

\textbf{Weighted voting.}
Using weights to expedite consensus has a long history. For example, back in the 80s, weighted voting differentiated the influence of servers by assigning them weights in the voting process~\cite{garcia1985assign, gifford1979weighted, jajodia1987dynamic}. However, these early approaches reach agreement only for single instances and work only in synchronous networks. In addition, with the rise of quorum replication, investigations into weighted voting and weighted partitions gained momentum~\cite{amir1998optimal, jimenez2003quorums, zhang2022escape}. 

Compared to HQC and majority-quorum consensus~\cite{kumar1991algorithm, zkweight}, weighted consensus addresses the trade-off between quorum size and the number of consensus rounds. Weighted consensus can reduce the quorum size while maintaining a two-round message-passing process to achieve consensus decisions.

\textbf{What's still missing?}
Despite decades of optimization in quorum replication, state-of-the-art approaches still struggle to meet the demands of modern computing, characterized by its large-scale and often heterogeneous nature. Notably, these approaches lack the following critical properties:

\begin{enumerate}[label=\textbf{P\arabic*}]  
    \item A universal weight assignment approach for conducting weighted consensus such that quorum replication can be reduced to any quorum size with flexible failure thresholds.
    \label{b:m1}

    \item A dynamic weight reassignment approach responsive to changing operating conditions such that the system maintains optimal performance.
    \label{b:m2}
\end{enumerate}

\ref{b:m1} enables a valuable tradeoff between fault tolerance and replication efficiency. Particularly, in large-scale systems, assuming a failure threshold $t$ less than the majority (i.e., $t < \ceil{\frac{n+1}{2}}$) is practically feasible. In addition, \ref{b:m2} empowers the system to dynamically prioritize best-performing nodes in weight reassignment, supporting the consensus process to maintain optimal performance.

Motivated by these challenges, \cab provides a precise definition with invariants of weighted consensus that allows for customizable fault tolerance. It dynamically reassigns weights by prioritizing more responsive nodes to maintain optimal performance. It offers an efficient yet straightforward implementation of weighted consensus based on Raft~\cite{ongaro2014search}. \cab provides a practical and effective solution for high-performance consensus algorithms, particularly for large-scale and heterogeneous systems.

\section{Weight Scheme}
\label{sec:weightscheme}

This section introduces weighted quorum and weight scheme, the most fundamental building blocks to achieve weighted consensus. Compared to conventional consensus (e.g., Paxos and Raft) that tolerates $f=\floor{\frac{n-1}{2}}$ failures by the simple majority rule, \textbf{weighted consensus offers a new tradeoff frontier to reduce the number of minimum failures} ($t$ where $1 \leq t \leq f$) \textbf{tolerated in exchange for higher performance}. For example, in a system of $10$ nodes, weighted consensus may set its failure threshold to $t=3$, whereas conventional consensus adheres to a fixed failure threshold of $f=4$.

A \textbf{weight scheme (WS)} is a sequence of weight values that designate the weight assigned to each node, while a \textbf{consensus threshold (CT)} determines the threshold required to reach an agreement. With a given $t$, we designate the $t+1$ nodes with the highest weights as \textbf{Cabinet Members}. In the following example, cabinet members are the $t+1$ nodes that have the $t+1$ highest weights ($w_1$ to $w_{t+1}$).
\begin{align}
\label{eq:wscheme}
\text{WS } = \underbrace{w_1, w_2, \ldots, w_t, w_{t+1}}_{\text{first $t{+}1$ values (\cab)}}, \underbrace{w_{t+2}, w_{t+3}, \ldots, w_{n-2}, w_{n-1}, w_n}_{\text{remaining $n{-}t{-}1$ values}}
\end{align}

Weight schemes should uphold the safety and liveness of weighted consensus algorithms.

\begin{description}
\item[Safety.] The WS and CT should underpin safety; i.e., no two correct nodes decide differently.

\item[Liveness.] The WS and CT should underpin liveness; i.e., a correct node eventually decides.

\end{description}

In conventional consensus, where every node weighs one, CT is simply the majority; i.e., $CT=\floor{\frac{n}{2}}+1$. CT cannot be set lower as it may allow more than one group of correct nodes to surpass CT and decide on different values~\cite{herlihy1986quorum,ongaro2014search}.

Similarly, not all WS and CT underpin both safety and liveness of weighted consensus. Figure~\ref{fig:three_ws} shows three weight schemes with $t=2$ and $n=7$, where $n_5$, $n_6$, and $n_7$ are cabinet members. For instance, in WS$_1$, weights are assigned based on node IDs with $CT=8$. However, WS$_1$ fails to ensure safety because correct nodes may reach different decisions. For example, $n_6$ and $n_7$ ($Sum(6,7)>8$) may decide on value $v$, while $n_2$, $n_3$, and $n_4$ ($Sum(2,3,4)>8$) may decide on another value, say $w$, violating safety.

Weighted voting suggests setting the consensus threshold to half of the total weights~\cite{jajodia1987dynamic}. Unfortunately, this alone does not suffice to guarantee both safety and liveness. For instance, in WS$_2$, weights grow exponentially, yielding $CT=\sum w_i/2=55555.5$. However, this scheme sabotages liveness. With $t=2$, the system should tolerate two failures, but in WS$_2$, even if only $n_7$ fails, the system halts because the combined weight of the remaining nodes falls below CT, preventing any agreement from being reached.

\begin{figure}[t]
\begin{tabular}{r|ccccccc|l}
    Nodes & $n_1$ & $n_2$ & $n_3$ & $n_4$ & $n_5$ & $n_6$ & $n_7$ & CT\\ \hline
    WS$_1$ & 1 & 2 & 3 & 4 & \textbf{5} & \textbf{6} & \textbf{7} & 8\\
    WS$_2$ & 1 & 10 & 10$^2$ & 10$^3$ & \textbf{10$^4$} & \textbf{10$^5$} & \textbf{10$^6$} & 55555.5\\
    \textbf{WS$_3$} & 2 & 3 & 4 & 6 & {\bf8} & {\bf10} & {\bf12} & 22.5 \\ \hline
\end{tabular}
\caption{Three weight schemes under $t=2$ and $n=7$, where $n_5$, $n_6$, and $n_7$ are cabinet members. WS$_1$ violates safety; WS$_2$ violates liveness; WS$_3$ can uphold both safety and liveness.}
\label{fig:three_ws}
\end{figure}

In \cab, we propose two invariants of weight schemes that can uphold both safety and liveness for weighted consensus. We also set the consensus threshold (CT) to half of the total weight. 
\begin{enumerate}[
label=\textbf{I\arabic*}]
    \item The sum of the $t+1$ highest weights (the weights of the cabinet members) is greater than the consensus threshold. \label{w:c1}
    
    \item The sum of the $t$ highest weights (the weights of the cabinet members excluding the lowest weight) is less than the consensus threshold. \label{w:c2}
\end{enumerate}
In other words, a weight scheme should adhere to Eq.~\ref{eq:2}.
\begin{align}
\label{eq:2}
    \sum_{i=1}^t w_i < \text{CT} = \sum_{i=1}^n \frac{w_i}{2} < \sum_{i=1}^{t+1} w_i
\end{align}

Invariants~\ref{w:c1} and~\ref{w:c2} serve as the foundation for the correctness of \textbf{weighted quorums} in terms of safety and liveness. With~\ref{w:c1}, when the group of cabinet members reaches an agreement, the remaining weights are not sufficient for any other agreement (see Lemma~\ref{lemma3.1}). With~\ref{w:c2}, when any $t$ nodes fail, the remaining weights are still sufficient to reach an agreement among alive nodes (see Lemma~\ref{lemma3.2}). Thus, when the cumulative weights surpass the consensus threshold, a system-wide decision is reached. Given that $t+1$ cabinet members always constitute the smallest-sized weight quorum, the fastest system-wide decision can be made once \cab members agree on the decision.

For example, WS$_3$ with $CT\!=\!22.5$ in Figure~\ref{fig:three_ws} is an eligible weight scheme that adheres to Eq.~\ref{eq:2} (i.e., $Sum(12,10)<22.5<Sum(12,10,8)$). Under normal operation, system-wide decisions can be made as fast as the cabinet members ($3$ nodes) have decided. Non-cabinet members cannot make a conflicting decision because their weights are less than the consensus threshold ($22.5>Sum(6,4,3,2)$). This weight scheme can tolerate at least $2$ failures as the total of weights excluding the highest two is greater than the consensus threshold ($Sum(8,6,4,3,2)>22.5$).

Next, we prove that weighted consensus achieves fast agreement where a system-wide decision can be made by the cabinet members agreeing on the decision. 

\begin{lemma} \label{lemma3.1}
The sum of the weights of all non-cabinet members is less than the consensus threshold.
\end{lemma}

\begin{proof}
We prove this lemma by contradiction. We denote the sum of the weights of the cabinet members is $W_{cab}$, and the sum of all weights is $W_{all}$. Given~\ref{w:c1}, $W_{cab} > W_{all}/2$. We claim that the sum of the weights of the remaining ${n-t-1}$ values is greater than the majority. As a result, $W_{all} - W_{cab} > W_{all}/2$; i.e., $W_{cab} < W_{all}/2$, which violates~\ref{w:c1}, Thus, this lemma holds.
\end{proof}

\begin{theorem}[Fast Agreement]
In every execution, if all cabinet members agree on a decision, then all non-faulty nodes agree on the same decision.
\end{theorem}

\begin{proof}
We prove this theorem by contradiction. We claim that all cabinet members agree on a decision $\mathcal{D}$, but at least one non-faulty node has decided on a conflicting decision, $\mathcal{D}_\diamond$.

Since $\mathcal{D}$ has been decided by cabinet members, the $t+1$ highest weights will not be voting for $\mathcal{D}_\diamond$. From Lemma~\ref{lemma3.1}, the sum of the weights of the remaining $n-t-1$ nodes (i.e., non-cabinet members) cannot surpass the consensus threshold. Thus, the decision for agreeing on $\mathcal{D}_\diamond$ cannot be made, which contradicts our claim. 
\end{proof}

Next, we prove that any weight scheme with a failure threshold of $t$ tolerates at least $t$ failures.

\begin{lemma} \label{lemma3.2}
The sum of the weights of any combination of $n-t$ nodes is greater than the consensus threshold.
\end{lemma}

\begin{proof}
We denote the nodes that are the cabinet members as set $N_{cab}$, where $|N_{cab}|=t+1$ with a total weight of $W_{cab}$, and the node that has the lowest weight in $N_{cab}$ as $n_{t+1}$ with weight $w_{t+1}$ (the ($t+1$)-th highest weight among all nodes). Given~\ref{w:c2}, following a similar way of proof, we know that the sum of weights of all non-cabinet members and $n_{t+1}$ (in a total of $n-t$ nodes) is greater than the consensus threshold. That is,
$$ \sum w_i + w_{t+1} > CT, \quad \text{for every } i \text{ where } n_i \notin N_{cab} $$

Since $n_{t+1}$ has the lowest weight in $N_{cab}$, every node that is not a cabinet member (i.e., $n_i \notin N_{cab}$) has a lower weight than $n_{t+1}$. Then, the above combination has the lowest weight of any combination of $n-t$ nodes. Therefore, the total weight of any combination of $n-t$ nodes must be greater than the consensus threshold. 
\end{proof}

\begin{theorem}[Fault Tolerance]
\label{theorem:3.2ft}
In every execution, weighted consensus tolerates at least $t$ failures.
\end{theorem}

\begin{proof}
In weighted consensus, an agreement can be reached when the sum of the weights of the non-faulty nodes is greater than the consensus threshold. Given Lemma~\ref{lemma3.2}, any $t$ failed nodes will not prevent an agreement from being achieved. Therefore, weighted consensus tolerates at least $t$ failures.
\end{proof}

\textbf{Flexible fault tolerance.} Weighted consensus offers a unique tradeoff between improving performance and tolerating failures. It can tolerate a minimum of $t$ failures (worst-case) and potentially a maximum of $n-t-1$ failures (best-case). The worst case arises when the top $t$ nodes with the highest weights experience failures (e.g., $n_6$ and $n_7$ fail in WS$_3$). In contrast, the best case occurs when all the cabinet members survive (e.g., $n_1$, $n_2$, $n_3$, and $n_4$ fail in  WS$_3$). In heterogeneous clusters, since strong nodes with higher weights are often less likely to fail than weak nodes, applications can reduce the quorum size and operate with fast agreement.

Weight schemes can be implemented in any ways as long as they satisfy \ref{w:c1} and \ref{w:c2}. In the following section, we introduce a simple and generalized implementation of weight schemes and present \cab in detail.

\section{The \cab Consensus Algorithm}
\label{sec:cabalgo}
We implemented \cab based on Raft~\cite{ongaro2014search}. Nodes in Raft operate in one of three states: \textit{leader}, \textit{follower}, and \textit{candidate}. In the normal case, there is one leader and the other nodes are followers. The leader issues \texttt{AppendEntriesRPCs} to followers, completing consensus as soon as a majority of the nodes respond to a single round of RPC calls. When the leader fails, followers become candidates and issue \texttt{RequestVoteRPCs} to start leader elections; the candidate that collects a majority of votes will become the new leader.

\cab implements weighted consensus on Raft by adding only two parameters to Raft's \texttt{AppendEntriesRPCs}. Raft's communication pattern and validation rules remain intact. In addition, \cab focuses on replication and does not change Raft's leader election mechanism. \cab preserves Raft's strong leadership concept: the leader, after being elected, computes a weight scheme based on a given failure threshold. It dynamically adjusts the weights of individual nodes based on their responsiveness, maintaining optimal performance throughout the consensus process.

\subsection{The Algorithm}
\cab first initializes a weight scheme according to the given failure threshold $t$. It uses geometric sequences to construct demanded weight schemes that satisfy Eq.~\ref{eq:2}. During the consensus process, \cab keeps reassigning higher weights to more responsive nodes, maintaining fast agreement and customized fault tolerance.

\subsubsection{\cab weight schemes}
\label{sec:cab:ws}
\cab implements weight schemes according to a given $t$ where $1 \leq t \leq \floor{\frac{n-1}{2}}$ by applying geometric sequences. When its common ratio $1<r<2$, a geometric sequence monotonically increases ($1<r$) with no value greater than the sum of its previous values ($r<2$). Initially, nodes are assigned weight values in descending order according to their node IDs, with higher IDs receiving lower initial weights.

\begin{tabular}{rcccccc}
    Weights: & $w_1$  & $w_2$  & $\ldots$ & $w_{n-2}$ & $w_{n-1}$ & $w_n$ \\
    Sequence: & $a_1 r^{n-1}$ & $a_1 r^{n-2}$ & $\ldots$ & $a_1 r^2$ & $a_1 r$ & $a_1$
\end{tabular}

\begin{figure}[t]
    \centering
    \begin{adjustbox}{width=\linewidth}
    \begin{tabular}{cc|cccccccccc}
         $t$ & $r$ & $w_1$ & $w_2$ & $w_3$ & $w_4$ & $w_5$ & $w_6$ & $w_7$ & $w_8$ & $w_9$ & $w_{10}$\\\hline
 
        $1$ & 1.40 & \textcolor{blue}{\bf20.7} & \textcolor{blue}{\bf14.8} & 10.5 & 7.5 & 5.4 & 3.8 & 2.7 & 2.0 & 1.4 & 1\\
        $2$ & 1.38 & \textcolor{blue}{\bf18.2} & \textcolor{blue}{\bf13.2} & \textcolor{blue}{\bf9.5} & 6.9 & 5.0 & 3.6 & 2.6 & 1.9 & 1.4 & 1 \\
        $3$ & 1.19 & \textcolor{blue}{\bf4.8} & \textcolor{blue}{\bf4.0} & \textcolor{blue}{\bf3.4} & \textcolor{blue}{\bf2.8} & 2.4 & 2.0 & 1.7 & 1.4 & 1.2 & 1 \\
        $4$ & 1.08 & \textcolor{blue}{\bf2.0} & \textcolor{blue}{\bf1.9} & \textcolor{blue}{\bf1.7} & \textcolor{blue}{\bf1.6} & \textcolor{blue}{\bf1.5} & 1.4 & 1.3 & 1.2 & 1.1 & 1 
    \end{tabular}
    \end{adjustbox}
    \caption{Cabinet weight schemes with different customized failure thresholds in a $n=10$ system. The weights of cabinet members in each scheme are in blue font.}
    \label{fig:exweights}
\end{figure}

To make the sequence eligible for the given $t$, the relation of weight values must be in accordance with Eq.~\ref{eq:2}, where the sum of the $t$ largest weights should be less than the consensus threshold while the sum of the $t+1$ largest weights should be greater than the consensus threshold (shown in Eq.~\ref{eq:3}).
\begin{align} \label{eq:3}
\sum_{i=n-t}^{n-1} a_1 r^i < CT = \frac{1}{2}\sum_{i=0}^{n-1} a_1 r^i < \sum_{i=n-t-1}^{n-1} a_1 r^i
\end{align}
Thus, the relation of $r$, $n$, and $t$ must satisfy Eq.~\ref{eq:4}.
\begin{align} \label{eq:4}
r^{n-t-1} < \frac{1}{2} (r^n +1) < r^{n-t}
\end{align}
Since the base value $a_1$ does not affect the progression of the sequence, for simplicity, we choose $a_1 = 1$. Figure~\ref{fig:exweights} shows examples of geometric sequences being used as weight schemes in a system of $n=10$ under $t=1, 2, 3, 4$. By changing the ratio $r$, the geometric sequence adapts to generate eligible weight schemes that can be used to achieve weighted consensus.

After the initial weight assignment, the weight of each node may change dynamically according to their responsiveness during the consensus process. It is worth noting that the leader redistributes the initial weights among nodes without generating any new weights. So far, \cab's weight scheme satisfies Property~\ref{b:m1}, providing a universal approach for quorum replication that underpins safety and liveness under customized failure thresholds. Next, we introduce the complete fast consensus algorithm with weight reassignment that satisfies Property~\ref{b:m2}.

\subsubsection{Fast consensus with weight reassignment}
\input{AlgoAugmentedRPC}

\cab adopts Raft's AppendEntries RPCs and \textit{adds only two parameters} to achieve the dynamic weight reassignment. The workflow with the parameters is described in Algorithm~\textsc{Cabinet-Consensus}. \cab introduces \textit{weight clock}, denoted by \texttt{wclock}, and \textit{weight value}, denoted by \texttt{$w_i$} to Raft's original AppendEntries RPCs (Line~\ref{algo:RPCwclock} and~\ref{algo:RPCweight}). The weight clock is simply a logical clock of the round of consensus instances. It is worth noting that \cab does not intervene in the original consensus tasks or other parameters defined by the Raft protocol, which upholds Raft's simplicity and correctness.

Initially, the leader calculates a weight scheme according to the given failure threshold $t$ and assigns nodes their initial weights (Line~\ref{algo:initweights}); it always assigns itself the highest weight ($w_{\lambda}$). After that, the weighted consensus starts with the leader retrieving each node's weight from the \textsc{GetWeight} function. Then, the leader issues the RPCs to update followers' weights and complete regular Raft's consensus tasks. Once the call completes, the leader adds $(weight$, $node)$ to the queue, $wQ$ (Line~\ref{algo:augstart} to~\ref{algo:augend}). Thus, the top $x$ pairs in $wQ$ represent the first $x$ replying followers in the weight clock $wclock$.

The leader keeps accumulating the weights of the replying nodes until they surpass the consensus threshold. The \textsc{UpdateWgt} function assigns the weights from the highest to the lowest on the weight scheme ($ws$) to nodes based on their orders dequeued from $wQ$. For example, the first node dequeued from $wQ$ will be the first to execute the \textsc{UpdateWgt} function and is assigned the second highest weight (the leader always takes the highest). Since $wQ$ is First in, First out, if node $n_i$ has replied faster than node $n_j$, then $n_i$ will be assigned a higher weight than $n_j$. Therefore, \textbf{in each weight clock, a node's weight value is dynamically adjusted based on its responsiveness.}  The leader assigns a higher weight value to a node in \texttt{wclock} $k+1$ if it received its reply earlier than others in \texttt{wclock} $k$. Thus, the leader and the top $t$ nodes whose replies are enqueued in $wQ$ in \texttt{wclock} $k$ become the cabinet members in \texttt{wclock} $k+1$.

When the accumulated weights (including the leader) exceed the consensus threshold, the weighted consensus is reached (Line~\ref{algo:consreached}). Then, the leader updates the weights for the remaining nodes (Line~\ref{algo:remaining}) whose weights were not previously updated at Line~\ref{algo:updwgtcab}. Note that the remaining nodes are not cabinet members. All these remaining nodes are assigned lower weights compared to the nodes that got their weights updated at Line~\ref{algo:updwgtcab}. 

\begin{figure}
    \centering
    \includegraphics[width=\linewidth]{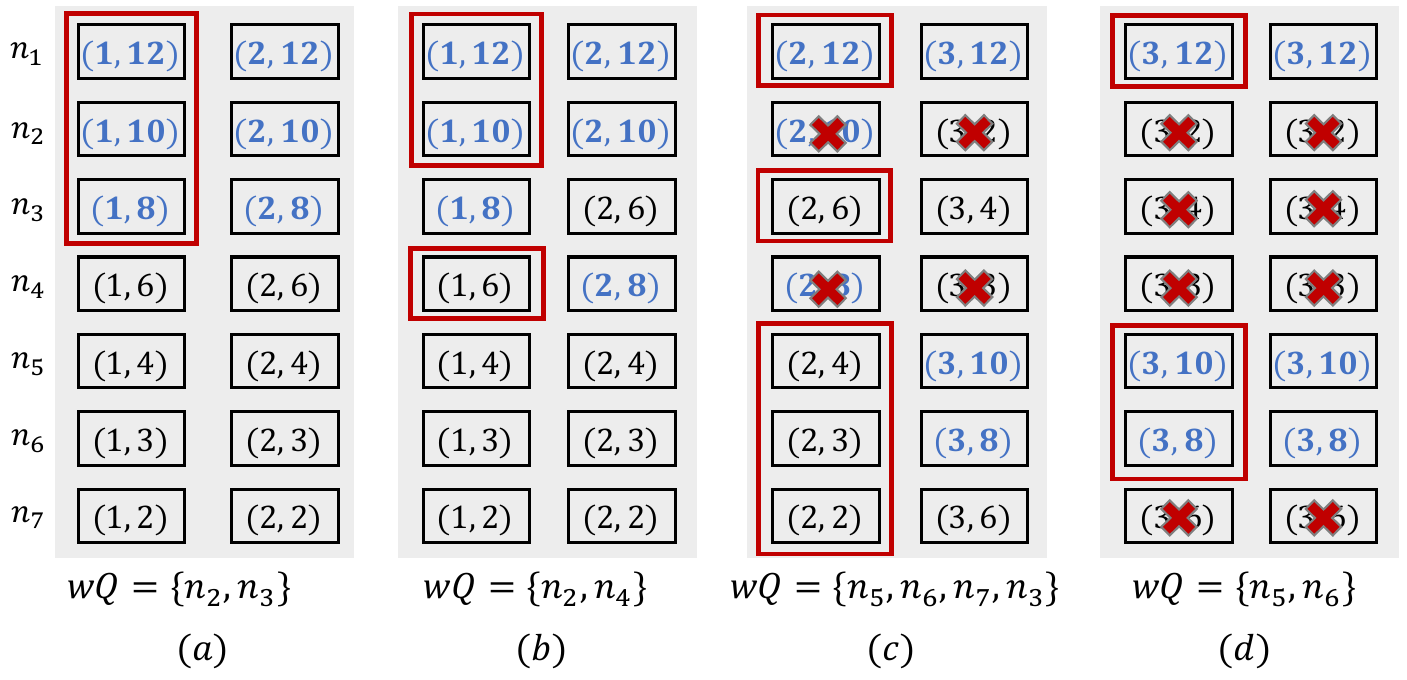}
    \caption{Possible weight distribution when $n=7$ and $n_1$ is the leader. The leader dynamically rearranges node weights $(wclock, w_i)$ based on the responsiveness received in $wQ$.}
    \label{fig:examples}
\end{figure}

\textbf{More examples.} Figure~\ref{fig:examples} shows examples of weight rearrangement in a \cab system of $n=7$ nodes where $t=2$. We use the weight scheme of the example introduced in \S\ref{sec:weightscheme} with a $CT=22.5$.

\begin{enumerate}[leftmargin=*, label=\textbf{(\alph*)}]
    \item When $wclock=1$, the cabinet members are $n_1$, $n_2$, and $n_3$ and have the $t+1$ highest weights among all nodes. If the cabinet members reply to the leader faster than the other nodes by the end of this execution, they remain to be cabinet members for the next execution (i.e., $wclock=2$). \label{ex:a}

    \item If $n_3$, as a cabinet member, fails to reply to the leader faster than $n_4$ in $wclock=1$, it will be excluded from the cabinet members with its weight assigned to $n_4$ in the next execution ($wclock=2$). \label{ex:b}

    \item Continuing from \ref{ex:b}, if $n_2$ and $n_4$ ($t$ cabinet members) crash, the leader can still coordinate consensus by collecting replies from the other nodes ($Sum(12,6,4,3,2) > CT$). Based on the reply order in $wQ$, $n_5$ and $n_6$ will become the cabinet members in the subsequent execution. \label{ex:c}
    
    \item Continuing from \ref{ex:c}, if all non-cabinet members fail, the consensus process will not be affected. Although the number of failed nodes ($f=4$) has exceeded the failure threshold ($t=2$), consensus can still be reached by cabinet members. In this case, \cab achieves a higher fault tolerance than conventional consensus approaches, which tolerate up to $f=3$ failures.\label{ex:d} 
\end{enumerate}

\textbf{Write and read.} 
In \cab, each node is required to store the consensus result along with the weight assigned to that particular consensus decision. Unlike conventional consensus where every node weighs the same, weighted consensus is no longer solely based on counting the number of nodes; therefore, it becomes necessary to explicitly indicate the node's weight associated with the corresponding consensus result. 

Once a client request is committed by the system, clients can read the consensus results. Read operations from clients should accumulate the stored weight of replies from nodes. Clients confirm that a proposed request is succeeded when the accumulated weights surpass the consensus threshold.

\textbf{Discussion.} By adding only two parameters to Raft's original RPCs, \cab efficiently implements weighted consensus with dynamic weight reassignment. Notably, the processes of weight reassignment and replication consensus are atomic, occurring within the same consensus instance and in the same RPC. This means that if the consensus to commit a value is successful, the weight update is applied; otherwise, neither operation succeeds.

With dynamic weight reassignments, \cab can seamlessly adapt to changing operating conditions. Compared to approaches that predefine voting or non-voting nodes, Cabinet eliminates the need for this rigid distinction. In a consensus round, nodes included in the quorum size are Cabinet members, which can be considered as the “voting” nodes, while the remaining excluded ones can be considered as “non-voting” nodes. Particularly, in heterogeneous clusters, where nodes possess varying capabilities, \cab's ability to assign weights to individual nodes allows for a more accurate representation of their influence in the consensus process. Strong nodes in these clusters will consistently maintain higher weights, solidifying their position as influential (cabinet) members.

In addition, \cab maintains the understandability of Raft; it preserves Raft's workflow and adheres to the concept of strong leadership, ensuring that the fundamentals of Raft remain intact. \cab's implementation of weighted consensus can seamlessly integrate within the existing Raft framework without altering its overall structure and clarity.

\subsubsection{Leadership change}
\label{sec:cab:le}

\cab adopts Raft's leader election mechanism without incorporating weighted consensus. Given the infrequency of leader elections and to maintain simplicity, \cab applies weighted consensus exclusively to replication. To comply with the quorum size of weighted consensus, \cab adjusts Raft's leader election mechanism by setting the quorum size to $n-t$; i.e., a candidate must collect votes from $n-t$ nodes to become the new leader, akin to Raft's majority quorum requiring $n-f$ votes. This adjustment ensures that an elected leader possesses the most up-to-date log as the minimum quorum size in replication is $t+1$. 

Furthermore, since \cab's weight reassignment prioritizes log responsiveness, in heterogeneous systems, the node with the most up-to-date log is one of the cabinet members. Although it can not be guaranteed that cabinet members are all strong nodes under complex networks (e.g., strong nodes may experience high delays and become less responsive), it increases the likelihood of a new leader being a strong node. We discuss and prove the correctness of this modification for Raft's leader election in \S\ref{sec:subsec:correctness}.

\subsubsection{Reconfiguration of failure thresholds}
\label{subsubsec:reconfig}
In addition to Raft's reconfiguration method that changes the cluster size, \cab introduces a lightweight reconfiguration for changing the failure threshold. First, the leader generates a new configuration, denoted by $C'$, with a new WS and CT that adhere to Eq.~\ref{eq:4}. It then starts to coordinate the consensus of changing the system configuration to $C'$. It first broadcasts $C'$ and waits for consensus to be reached under the new WS. Note that no replication will be conducted during the transition from the old configuration to the new configuration. Once consensus of $C'$ is achieved under the new WS, the leader transitions to operating under the new failure threshold, similar to Raft's reconfiguration procedure. \cab's reconfiguration of failure thresholds offers applications an adaptive way to tradeoff between absolute fault tolerance and system performance in evolving operating environments. 

\subsection{Flexible Fault Tolerance}
\label{sec:subsec:liquefiedft}
\cab obtains flexible fault tolerance in two distinct dimensions. The first dimension is the customized failure threshold ($t$), which empowers applications to define their own failure thresholds based on their preferences and user experience. The customization becomes especially valuable when large-scale applications that may not require tolerance for $\floor{\frac{n-1}{2}}$ simultaneous failures but prioritize higher performance. 

The second dimension lies in its flexible fault tolerance capabilities in any established weight scheme, allowing for the tolerance of at least $t$ and up to $n-t-1$ physical node failures. In the worst-case scenario, \cab can tolerate $t$ failures, which occurs when the top $t$ nodes with the highest weights experience failures. However, in the best-case scenario where all cabinet members remain operational, the system demonstrates its resilience by tolerating up to $n-t-1$ physical node failures. This surpasses the fault tolerance limits of conventional consensus approaches, as exemplified in the scenarios presented in Examples~\ref{ex:c} and~\ref{ex:d}. The ability of \cab to dynamically adjust weights and adapt to changing node conditions empowers it to provide a fault tolerance mechanism that outperforms conventional consensus algorithms in accommodating both extreme and more favorable failure scenarios.

The two-dimensional liquefaction of fault tolerance is especially advantageous in heterogeneous clusters. Strong nodes, assigned with high weights, have a higher likelihood of resilience, while weaker nodes, assigned with lower weights, are more susceptible to failure. Consequently, \cab will operate closer to the best-case scenarios, where failures are less likely among cabinet members, allowing it to exceed the expected $t$ tolerance in practice. 

\subsection{Correctness Argument}
\label{sec:subsec:correctness}
We now discuss the correctness of \cab in replication and in leader election. Since \cab is implemented based on Raft, we show the added weight mechanism does not affect correctness; i.e., \cab maintains Raft's safety and liveness properties. 

\subsubsection{Correctness}
\label{subsec:ap:correctargument}
\cab inherits Raft's \textsc{Integrity} property; i.e., every correct node decides at most one value. The proposed weighted consensus mechanism does not change the rule of nodes accepting values but changes only the quorum size. We argue \cab's correctness by showing that \cab maintains Raft's validity in the consensus process.

\begin{theorem}[Validity]
If some correct node decides a value $v$, then $v$ is the initial value of some node.
\end{theorem}

\begin{proof}
\cab's requires a minimal size of 2 cabinet members when $t=1$ (the lower bound). Thus, at least one cabinet member must be correct under any choice of $t$. If a value $v$ is committed, at least one correct cabinet member has agreed on $v$. According to \cab's message passing, the leader that leads the consensus process of $v$ must have proposed $v$ to at least one correct node.
\end{proof}

Before we argue \cab's safety and liveness. We reiterate that \cab adopts Raft's leader election mechanism only with a modification of the election quorum size (see \S\ref{sec:cab:le}). That is, instead of collecting votes from a majority of nodes, a candidate must collect votes from $n-t$ nodes, which is in accordance of \cab's customized failure threshold. \cab does not impose further changes to Raft's original specifications. Since $t$ is typically set to a smaller value than the majority threshold (i.e., $1 \leq t \leq f=\floor{\frac{n-1}{2}}$), \cab requires a larger size of the election quorum than Raft. However, compared to replication, leader election is infrequent in leader-based consensus algorithms, the impact of the enlarged election quorum size on overall system performance can be neglected.

To understand election safety in \cab, we first show that the modified quorum size still enforces the election of the most up-to-date nodes as a new leader.

\marginpar{\hspace{0.5em}\textcolor{gray!50}{\vrule height 12cm width 2.5pt}} 

\begin{lemma} \label{lemma:leuptodate}
By requiring an election quorum of size $n-t$, an elected leader in \cab must have the most up-to-date log.
\end{lemma}

\begin{proof}
We partition the $n$ nodes into two groups: $|G_1|=t+1$ and $|G_2|=n-t-1$, where $|G_1|+|G_2|=n$. Consider a consensus instance replicating a value $v$. In \cab, the minimum quorum size is $t+1$, which corresponds to all cabinet members (represented by $G_1$). Nodes in $G_1$ possess the most up-to-date logs, while those in $G_2$ lag behind.

In Raft, a node only votes for a candidate that is as least as up-to-date as itself. Thus, a lag-behind candidate can collect at most $n-t-1$ votes (i.e., from nodes in $G_2$). Since \cab's election quorum size is $n-t$, any candidate that does not possess the up-to-date log cannot be elected as a new leader. 

When the number of nodes that replicate $v$ is greater than $t+1$, then stall nodes are less than $n-t-1$. Consequently, a stall candidate will receive fewer votes than $n-t-1$, which is a worse case than the aforementioned case. The stall candidate cannot be elected. Therefore, \cab ensures that an elected leader must have the most up-to-date log.
\end{proof}

\begin{lemma} \label{lemma:lecsafety}
\cab maintains Raft's election safety; that is, at most one leader can be elected in a given term.
\end{lemma}

\begin{proof}
In Raft, each node only votes once in a given term and votes for a candidate that is as least as up-to-date as itself. Thus, there can only be one elected candidate that has the most up-to-date log. When votes are split among eligible candidates, Raft simply waits for new timeouts that increase the current term and repeats the election process until a new leader is elected.

\cab adopts all the above policies without any changes. Since \cab's failure threshold $t$ is always less than $f$ (i.e., $1 \leq t \leq f=\floor{\frac{n-1}{2}}$; see~\S\ref{sec:cab:ws}), at most one candidate can collect $n-t$ votes in a term. Thus, \cab cannot elect more than one leader in a given term, thereby inheriting Raft's election safety.
\end{proof}

Note that the split vote problem in Raft can still occur in \cab, as their election criteria are identical except for the quorum size. When more than one eligible follower transitions to be a candidate, they may collect partial votes where no one can collect sufficient votes (i.e., $n-f$ votes) and thus no one wins the election. The election will repeat when a new time out is triggered from other followers or current candidates until a leader is elected. Note that split votes are extremely rare when timers are randomized. For more details on setting randomized timeouts, see the discussion of leader election in Raft's specification~\cite{ongaro2014search}.

With election safety, we now discuss \cab's safety in the entire consensus process (i.e., both replication and election). In \S\ref{sec:weightscheme}, we have shown the unique properties of fast agreement and fault tolerance of weighted consensus in one consensus execution. Now we sketch the proof for \cab's safety.

\begin{theorem}[Safety]
No two correct nodes decide differently.
\end{theorem}

\begin{proof}(Sketch)
From Lemma~\ref{lemma:leuptodate} and~\ref{lemma:lecsafety}, \cab inherits the correctness of Raft's leader election mechanism; thus, the \textit{election safety} and \textit{leadership completeness} introduced in Raft are maintained. Once a leader is elected, the replication process commences with nodes assigned distinct weights in each consensus execution. 

\marginpar{\hspace{0.5cm}\textcolor{gray!50}{\vrule height 21.6cm width 2.5pt}} 

Raft's leader election mechanism guarantees that a new leader comes from the most up-to-date nodes in the system. As weight clocks monotonically increase in \cab's AppendEntries RPCs, a new leader must have the highest weight clock of the latest system-wide committed value. Consequently, the new leader will not overwrite previous weight clocks, ensuring that weight distinctness is maintained throughout the consensus process.

Given Lemma~\ref{lemma3.1}, in each weight clock, non-cabinet members cannot decide on a different value when cabinet members have decided. Therefore, \cab prohibits correct nodes from deciding differently.
\end{proof}

\begin{theorem}[Liveness] \label{theorem:liveness}
A correct node eventually decides.
\end{theorem}

\begin{proof}
We show that \cab's liveness property is equivalent to Raft's liveness property. We discuss the equivalence in leadership changes and replication. 

By changing the election quorum size to $n-f$, \cab guarantees that there are always sufficient votes for a single node to become a leader. Since there are at most $f$ failures, the remaining nodes (i.e., $n-f$ nodes) are all correct nodes. When a candidate's log is up-to-date and in the highest term, correct nodes will vote for the candidate and from an election quorum size of $n-f$.

By Lemma~\ref{lemma:lecsafety}, \cab is able to reuse Raft's leader election mechanism. Thus, the liveness in leadership changes in \cab is equivalent to that in Raft. In addition, by Theorem~\ref{theorem:3.2ft}, in each consensus execution, a \cab system with $n$ nodes can at least tolerate $t$ failures ($t \leq f=\floor{\frac{n-1}{2}}$, where $f$ is the failure threshold in Raft). Thus, the combination of leader election and replication in \cab can at least tolerate $t$ failures. Consequently, \cab upholds the same liveness property as Raft.
\end{proof}

Theorem~\ref{theorem:liveness} shows the lower bound liveness of \cab. As discussed in \S\ref{sec:subsec:liquefiedft}, \cab may exhibit stronger liveness than Raft by operating under more than $f=\floor{\frac{n-1}{2}}$ node failures as long as the total weights of remaining nodes exceed its consensus threshold.

\begin{figure}[t]
    \centering
    \begin{tabular}{c|ccccc|c}
    nodes & 1 & 2 & 3 &4 &5 & $w_{i}$\\ \hline
    $n_1$ & x & x & x &x &x & $w_{1}$\\
    $n_2$ & x & x & x &x &x & $w_{2}$\\
    $n_3$ & x & x & x & & & $w_{3}$\\
    $n_4$ & x & x & x & & & $w_{4}$\\
    $n_5$ & x & x &  & & & $w_{5}$\\
    \end{tabular}
    \caption{A possible log snapshot of \cab cluster of $5$ nodes with $t=2$. $n_1$ and $n_2$ are cabinet members in Instance 5.}
    \label{fig:ap:snapshot}
\end{figure}

\subsubsection{Leader properties}
\label{subsec:ap:leadership}
Expanding on the earlier discussion regarding the properties of new leaders, \cab's weight reassignment strategy prioritizes log responsiveness. In each consensus instance, the nodes that have the most up-to-date logs are the cabinet members; thus, a leader will be elected from the cabinet members of the latest consensus instance of replication.

\marginpar{\hspace{0.5em}\textcolor{gray!50}{\vrule height 17.8cm width 2.5pt}} 

For example, consider a \cab system with $n=5$ and $t=1$, illustrated by the snapshot of node logs depicted in Figure~\ref{fig:ap:snapshot}, where ``x'' denotes nodes that have replicated the entry coordinated by the leader. Assume node $n_1$ is the leader and fails. Since $t=1$, the replication quorum (cabinet members) $t+1$ is 2. In Figure~\ref{fig:ap:snapshot}, $n_1$ and $n_2$ are cabinet members and have the most up-to-date logs. Even through $n_2$, $n_3$, $n_4$, and $n_5$ are correct nodes, by weighted consensus, $n_3$, $n_4$, and $n_5$ are allowed to fall behind.

In this election, only $n_2$ is eligible to become the new leader, given that it has the most up-to-date log and can receive votes from all the remaining nodes (including itself). The other remaining nodes may also start an election, but no one can win as $n_2$ will only be voting for itself. Therefore, the elected leader must emerge from the cabinet members during the latest replication phase before the election occurs.

In heterogeneous systems, it is common for a strong node to possess the most up-to-date log and be selected as a new leader. While this outcome is often expected, it cannot be guaranteed. For instance, in complex networks, strong nodes may encounter high delays, rendering them less responsive. In such scenarios, during an ongoing consensus instance, strong nodes might be assigned lower weights. Consequently, during an election at such moments, a weak node, when having the most up-to-date log, could potentially be elected as the new leader.

Nevertheless, in typical scenarios where strong nodes are more responsive due to their processing capability, \cab can increase the likelihood of a new leader elected from a pool of strong nodes. This emphasis on selecting leaders from nodes with the most up-to-date logs contributes to the overall robustness and performance of the system.

\marginpar{\hspace{0.5cm}\textcolor{gray!50}{\vrule height 9.5cm width 2.5pt}} 

\section{Evaluation}
\label{sec:evaluation}
We implemented Raft and \cab in Golang and evaluated their performance on a popular cloud platform~\cite{ccc}. All nodes were equipped with 2.40 GHz Intel Xeon (Skylake) processors, running Ubuntu 18.04.1 LTS. The TCP/IP bandwidth is $\approx 400$ Megabytes/s, with a raw network latency $<1$~ms. Our experiments are conducted in both homogeneous and heterogeneous clusters, where VMs are grouped into distinct \textit{zones} of different configurations. 

\textbf{Heterogeneous clusters:}
Heterogeneity across zones is defined by variations in the number of vCPUs, RAM, and disk space. For each scale $n=3, 5, 7, 11, 20, 50, 100$, the five configurations are evenly distributed among the VMs, where "\texttt{\textcolor{blue}{$\#x$}c-\textcolor{blue}{$\#y$}gb-\textcolor{blue}{$\#z$}}" represents ``$\#x$ vCPU, $\#y$GB RAM, and $\#z$GB Disk". 

\textbf{Homogeneous clusters:}
All VMs have the same configuration of Z3. The uniform configuration remains consistent across the evaluated cluster scale, ensuring that all VMs within the cluster have identical configurations.

\begin{tabular}{cc|ccccccc}
     Zone & Heterogeneity & 3 & 5 & 7 & 11 & 20 & 50 & 100\\
     \hline
     Z1 & 1c-7.5gb-56 & 1 & 1 & 2 & 2 & 4 & 10 & 20\\
     Z2 & 2c-15gb-92  &   & 1 & 1 & 2 & 4 & 10 & 20\\
     Z3 & 4c-15gb-164 & 1 & 1 & 1 & 2 & 4 & 10 & 20\\
     Z4 & 8c-30gb-308 &   & 1 & 1 & 2 & 4 & 10 & 20\\
     Z5 & 16c-60gb-596 & 1 & 1 & 2 & 3 & 4 & 10 & 20\\
\end{tabular}

We introduce the following notations to report the results:\\
\textbf{``n"} represents the cluster size; i.e., the number of VMs.\\
\textbf{``d''} represents the implemented network delays (see \S\ref{sec:eval:perfcomplexnetworks}).\\
\textbf{``b"} represents the batch size.\\
\textbf{``cab $f\textcolor{blue}{x}\%$''} represents that \cab's $t=\textcolor{blue}{x}\%$ of $n$; e.g., ``cab  $f10\%$'' under $n=50$ means \cab's $t=5$.

The rest of this section is organized as follows. It introduces the proposed benchmark framework for distributed consensus applications (\S\ref{sec:eval:benmarkfw}). It reports the performance of \cab with varying failure thresholds compared against Raft in both heterogeneous and homogeneous clusters (\S\ref{sec:eval:perfscalingcluster}), complex network conditions of differently distributed and dynamically changing network delays (\S\ref{sec:eval:perfcomplexnetworks}), and node failures under various crash scenarios (\S\ref{sec:eval:perfuderfailures}).

\subsection{\cab Benchmark Framework}
\label{sec:eval:benmarkfw}

We propose a benchmark framework tailored for distributed consensus applications, with a special focus on leader-based consensus algorithms. Our framework introduces a set of versatile and adaptable interfaces that seamlessly integrate existing benchmarks. In the context of distributed database applications, it has integrated the widely adopted YCSB~\cite{cooper2010benchmarking} and TPC-C~\cite{tpcc} workloads.

\begin{figure}[t]
    \centering
    \includegraphics[width=0.99\linewidth]{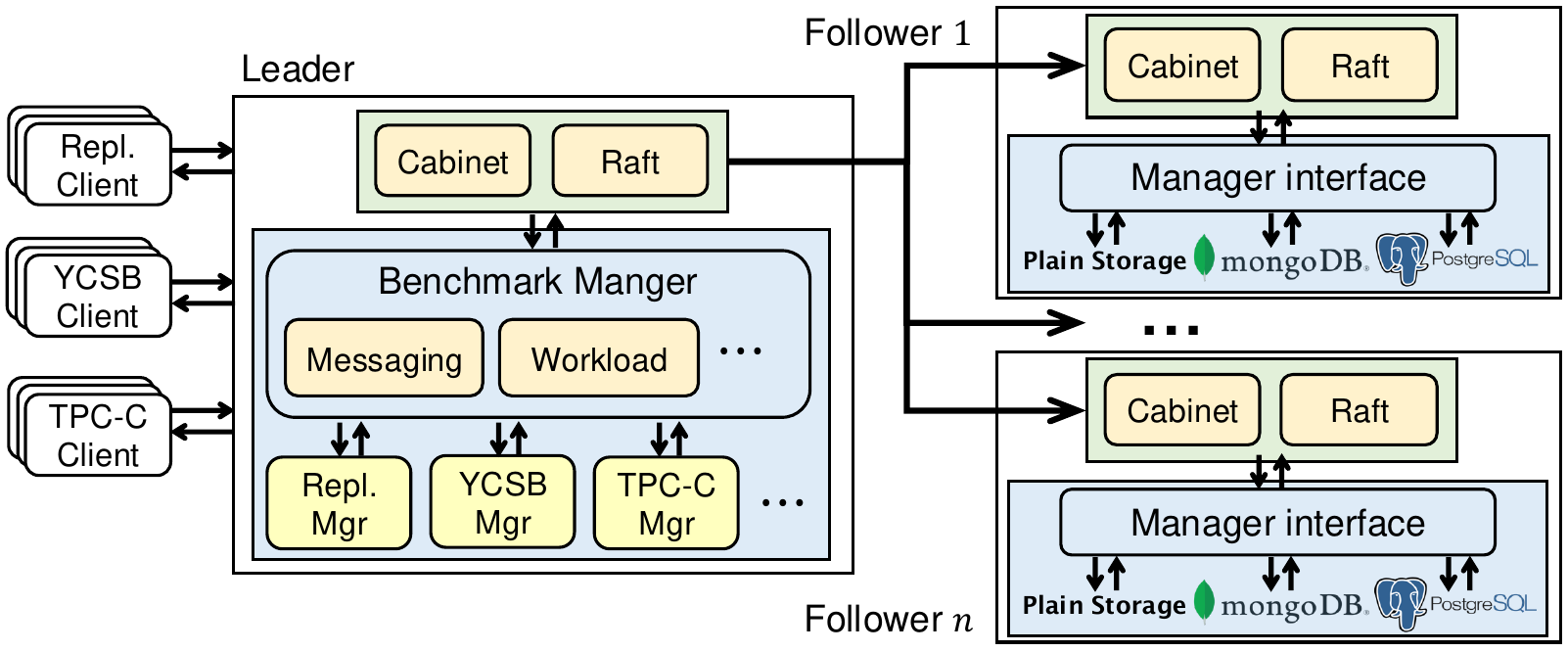}
    \caption{The architecture of \cab benchmark framework.}
    \label{fig:evalstructure}
\end{figure}

The architecture of our framework is depicted in Figure~\ref{fig:evalstructure}, providing a comprehensive overview of its key components and their interactions. The leader node takes charge of orchestrating and managing different integrated benchmarks. Each integrated benchmark is equipped with its dedicated manager. These managers act as control centers, enabling users to fine-tune and customize evaluation parameters for specific requirements, such as batch sizes, workload sizes, ratios of workload types, and combinations of transactions. In addition, the leader batches workload data proposed from clients and issues RPCs piggybacking the consensus metadata and the batched workload data to followers. Followers encompass distributed applications that execute the transmitted workload data. 

The proposed framework reports throughput and latency for the consensus process and is also able to report other criteria contained in the integrated benchmark. By leveraging our proposed framework, we can conduct evaluations under various workloads, either by creating custom workloads or by utilizing existing workloads provided by the integrated benchmarks. This flexibility empowers users to simulate real-world scenarios and accurately evaluate the performance of consensus algorithms in different contexts.

\begin{figure*}[t]
\minipage{0.5\textwidth}
    \centering
    \includegraphics[width=\linewidth]{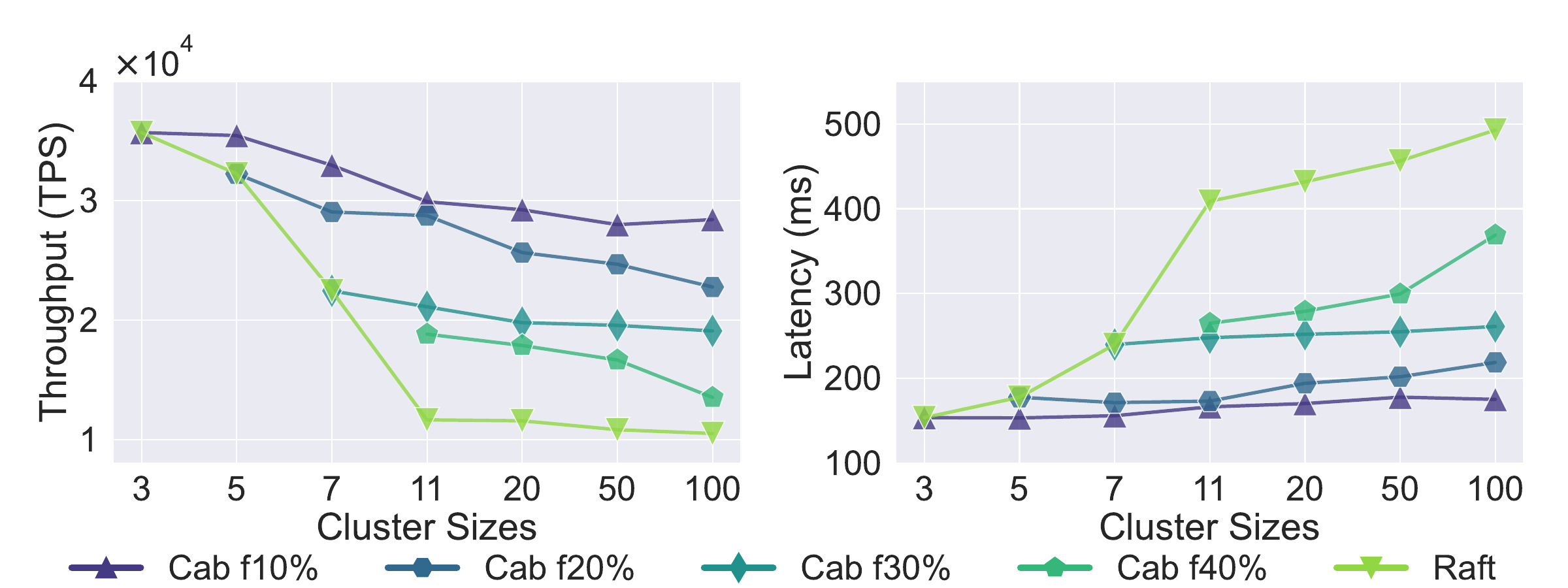}
    \subcaption{Heterogeneous cluster.}
    \label{fig:ycsbscalesd0hetero}
\endminipage \hfill
\minipage{0.5\textwidth}
    \centering
    \includegraphics[width=\linewidth]{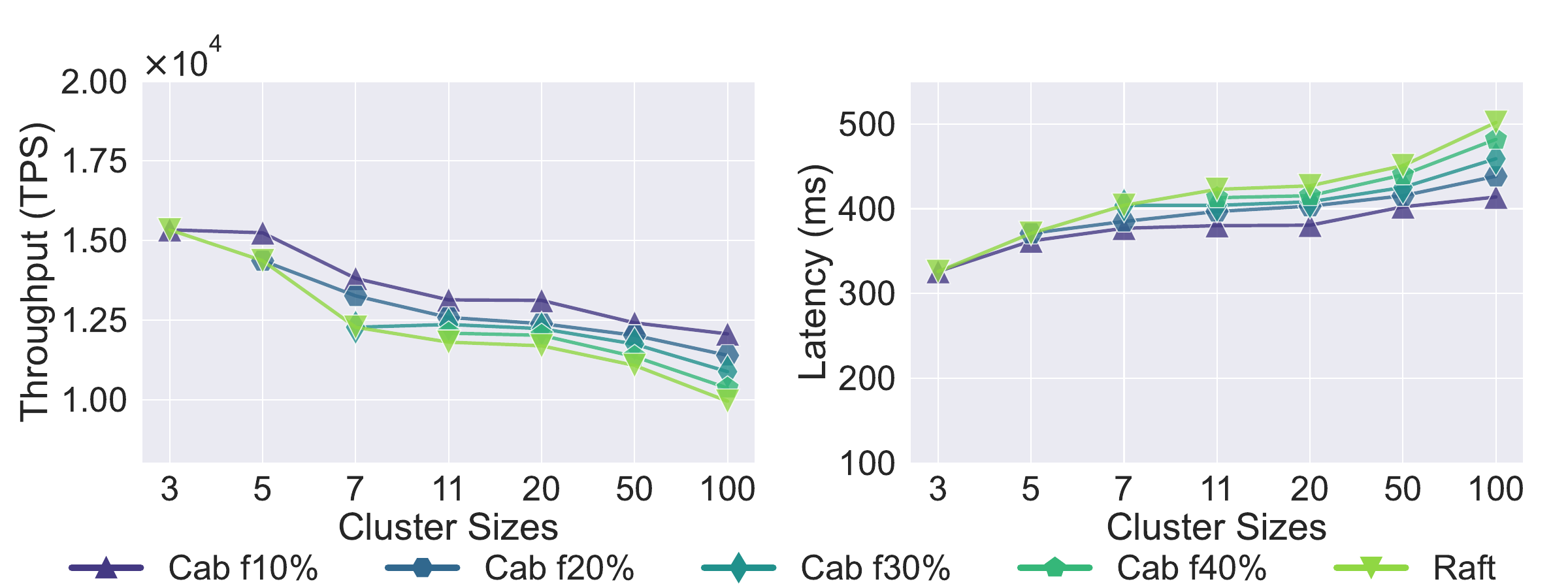}
    \subcaption{Homogeneous cluster.}
    \label{fig:ycsbscalesd0homo}
\endminipage
\caption{Performance of YCSB+MongoDB under increasing cluster sizes with $b=5k$, $d=0$ under Workload A.}
\label{fig:ycsbscales}
\end{figure*}

\begin{figure*}
\minipage{0.5\textwidth}
    \centering
    \includegraphics[width=\linewidth]{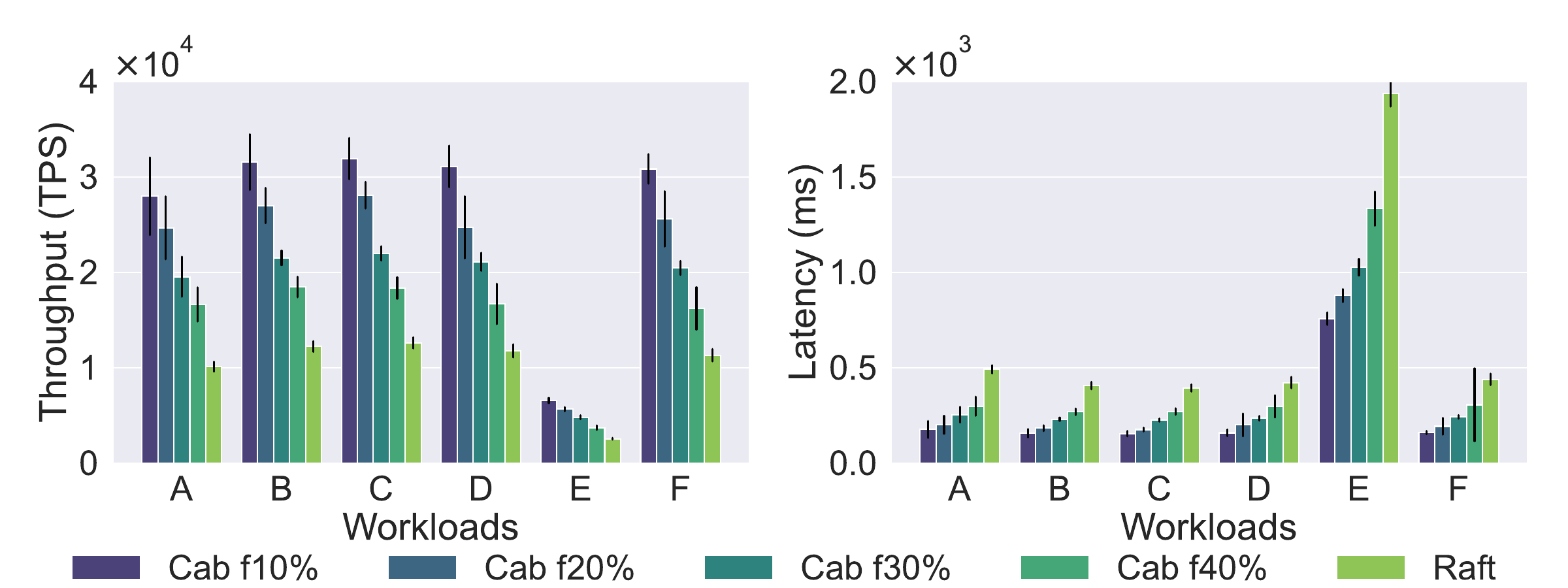}
    \subcaption{Heterogeneous cluster.}
    \label{fig:ycsb50d0hetero}
\endminipage \hfill
\minipage{0.5\textwidth}
    \centering
    \includegraphics[width=\linewidth]{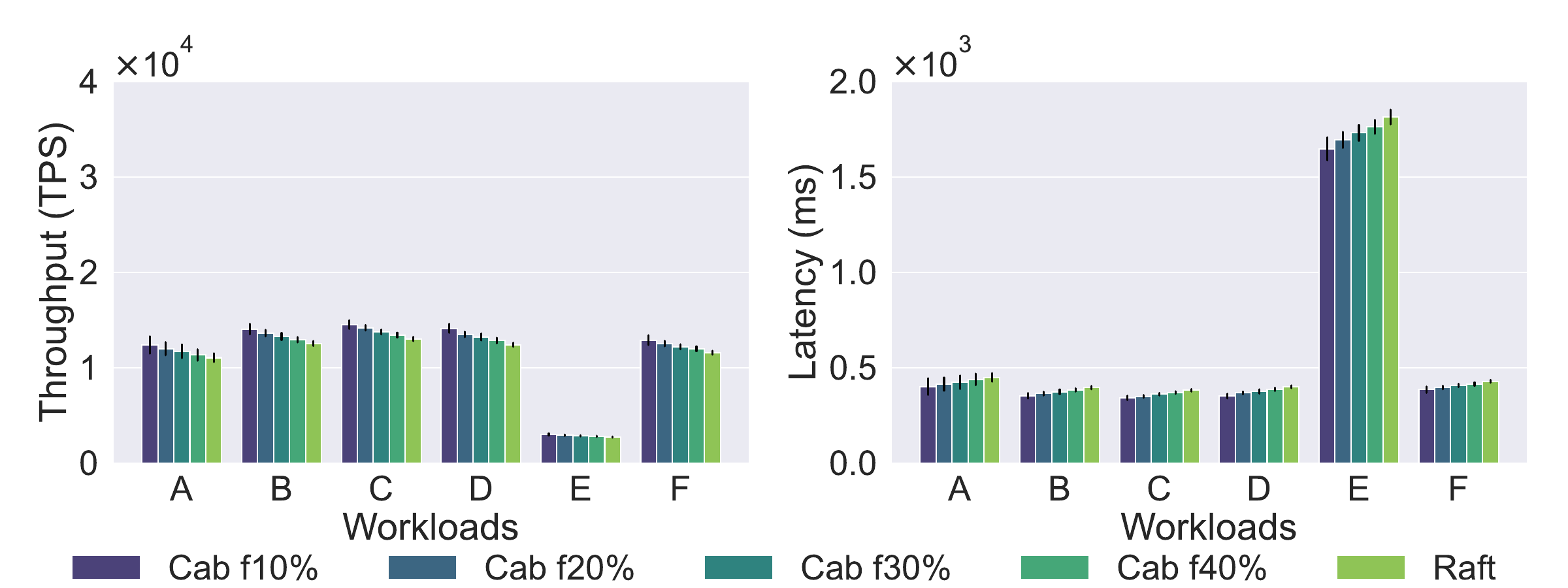}
    \subcaption{Homogeneous cluster.}
    \label{fig:ycsb50d0homo}
\endminipage
\caption{Performance of YCSB+MongoDB under $n=50$, $b=5k$, $d=0$ under all YCSB workloads.}
\label{fig:ycsbn50d0}
\end{figure*}

In this evaluation, we deployed the consensus module with \cab and Raft~\cite{ongaro2014search}, where \cab has four failure thresholds of $t=10\%n, 20\%n, 30\%n, 40\%n$ for each cluster size of $n=10, 20, 50, 100$. The applied databases on followers are MongoDB~\cite{mongodb}, which is a popular document-oriented NoSQL database, and PostgreSQL~\cite{postgresql}, which is a widely used relational database management system. In addition, our evaluation paired MongoDB with YCSB~\cite{cooper2010benchmarking} and PostgreSQL with TPC-C~\cite{tpcc} benchmarks. 

\textbf{YCSB+MongoDB:}
YCSB is often used to compare the relative performance of NoSQL database management systems. It consists of six standard workloads (from \textbf{A} to \textbf{F}), each with a specific distribution of four operation types: READ, UPDATE, SCAN, and INSERT (details presented in~\cite{onlineap}). In the \texttt{YCSB+MongoDB} evaluation, we conducted 10 runs of a specific workload, with each run involving the leader processing a total of 500,000 corresponding operations, where the batch size was set to $b=5,000 (5k)$. As a result, each run concluded after completing 100 rounds of processing.

\textbf{TPC-C+PostgreSQL:} TPC-C~\cite{tpcc} is one of the most widely used benchmarks for evaluating the performance and scalability of transactional databases for online transaction processing. It simulates a complete order-entry system, including a predefined mix of transactions performed by concurrent users. In the evaluation of \texttt{TPC-C+PostgreSQL}, a total of 50,000 transactions are processed by the leader in batches of size $b=2,000$ (2k). The transaction mix within each batch follows the same predefined ratio. Each follower instance of PostgreSQL is configured with 10 warehouses and 16 users. We conducted 10 independent runs for each experiment.

\begin{figure*}
\minipage{0.5\textwidth}
    \centering
    \includegraphics[width=\linewidth]{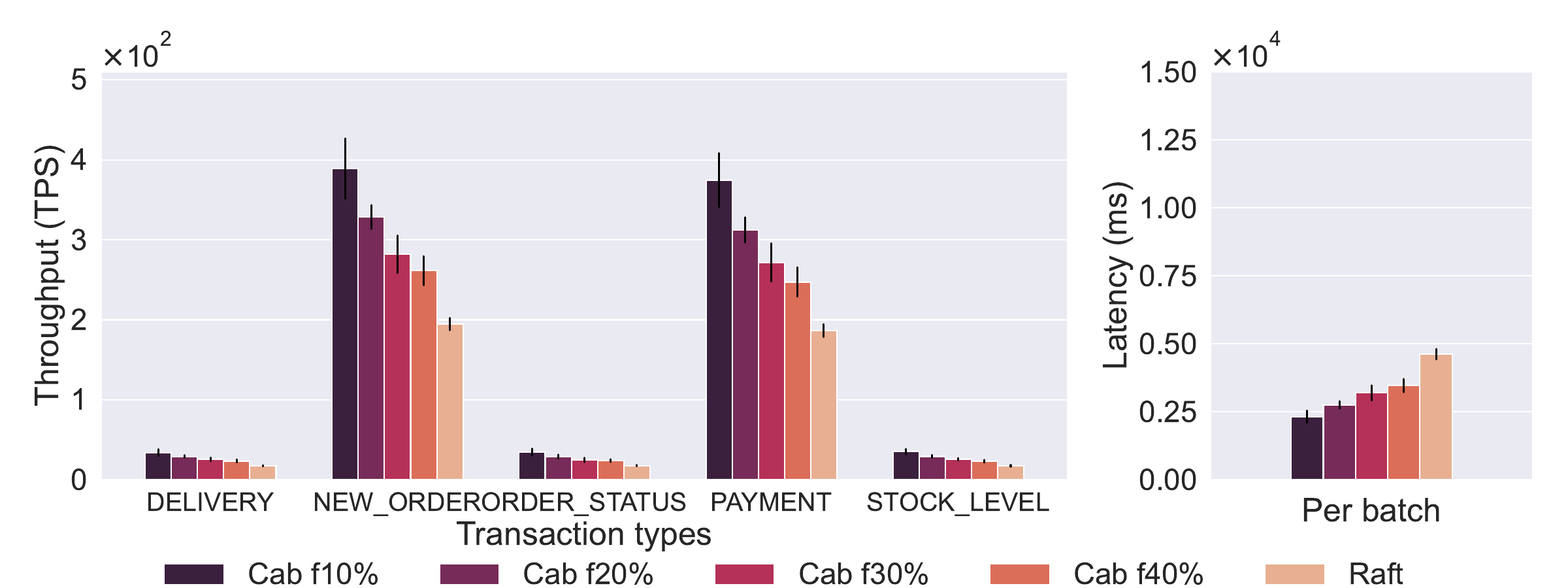}
    \subcaption{Heterogeneous cluster.}
    \label{fig:tpcc50d0hetero}
\endminipage \hfill
\minipage{0.5\textwidth}
    \centering
    \includegraphics[width=\linewidth]{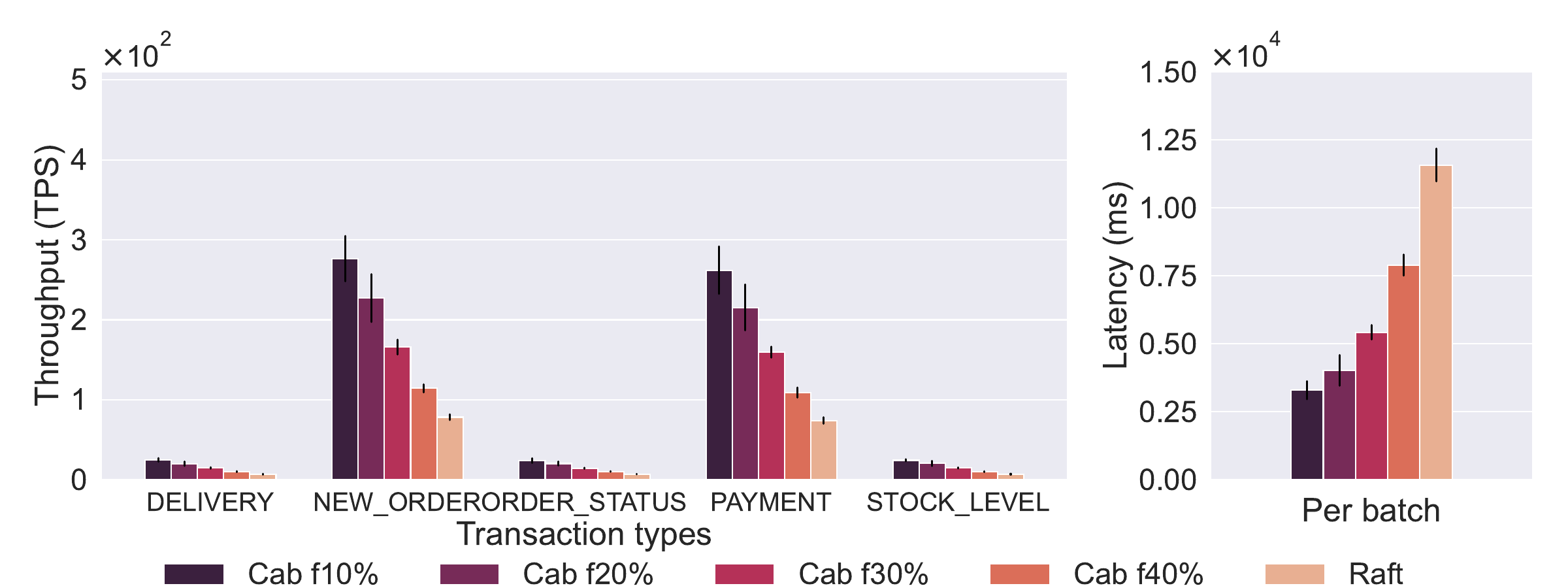}
    \subcaption{Homogeneous cluster.}
    \label{fig:tpcc50d0homo}
\endminipage
\caption{Performance of TPC-C+PostgreSQL under $n=50$, $b=2k$, $d=0$ under all TPC-C workloads.}
\label{fig:tpccn50d0}

\minipage{0.5\textwidth}
    \centering
    \includegraphics[width=\linewidth]{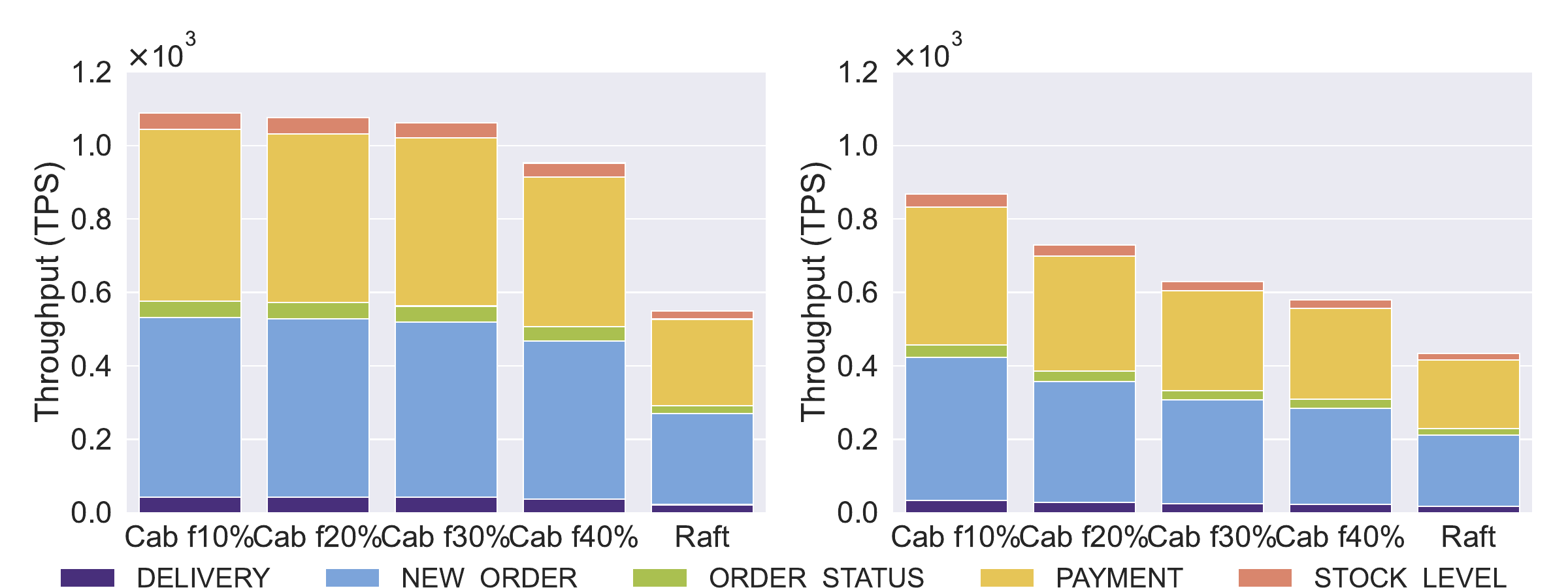}
    \subcaption{Heterogeneous cluster (left $n=11$, right $n=50$).}
    \label{fig:tpccscaleshetero}
\endminipage \hfill
\minipage{0.5\textwidth}
    \centering
    \includegraphics[width=\linewidth]{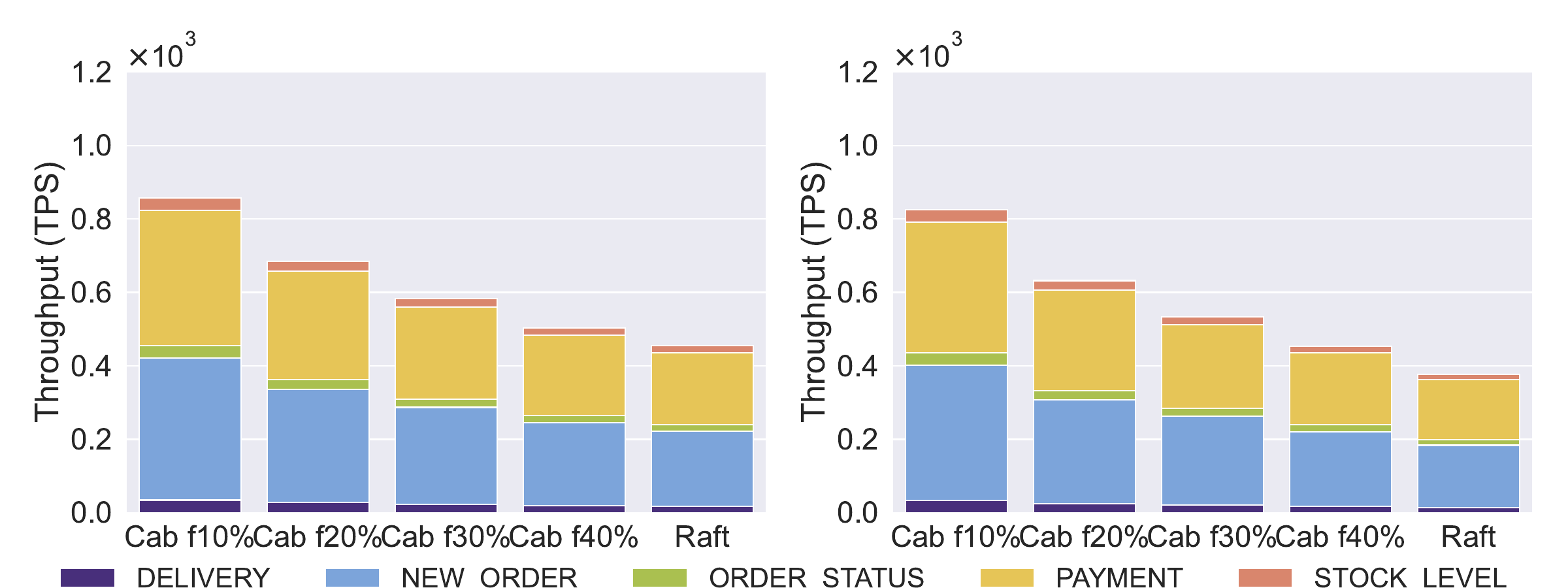}
    \subcaption{Homogeneous cluster (left $n=11$, right $n=50$).}
    \label{fig:tpccscaleshomo}
\endminipage
\caption{Performance of TPC-C+PostgreSQL under two cluster sizes ($n=11$, $50$) with $b=2k$, $d=0$ under all workloads.}
\label{fig:tpccscales}
\end{figure*}

\subsection{Performance under Scaling Clusters}
\label{sec:eval:perfscalingcluster}

We evaluated the performance of YCSB and TPC-C across different scales, where $n=3$, $5$, $7$, $11$, $20$, $50$, and $100$, using various workloads. Due to space limitations, we present the results of only Workload~A in Figure~\ref{fig:ycsbscales}. With $n=3$, \cab and Raft have the quorum size of $2$ with near identical performance. With $n=5$, \cab can choose $t=1$ or $t=2$, where $t=2$ is the same as Raft. When $n \geq 5$, \cab starts to show the advantage of weighted consensus by reducing the quorum size.

\begin{figure}[t]
    \centering
    \includegraphics[width=0.95\linewidth]{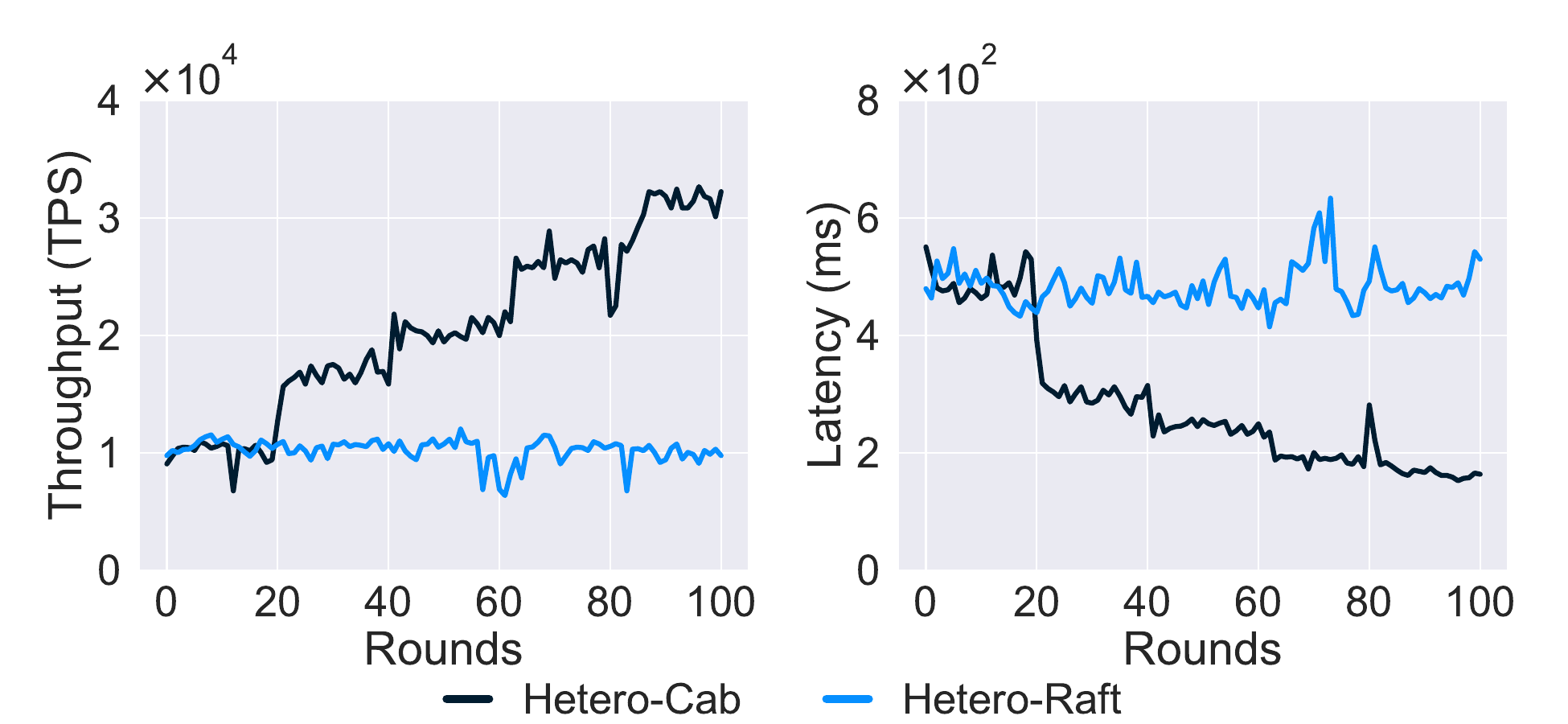}
    \caption{Throughput and latency in heterogeneous clusters of YCSB+MongoDB with $d=0, n=50, b=5k$ under Workload A, where $t {=} 25$ (initial), $20$, $15$, $10$, and $5$ for every $20$ rounds.}
    \label{fig:ycsbchangingt}
\end{figure}

Under a larger scale, compared to Raft, \cab witnesses a more pronounced advantage in the heterogeneous setting. It is worth noting that the performance loss when scaling up the system is minimal for both \cab and Raft. This is because both \cab and Raft achieve agreement in just one round of remote procedure calls, making them linearly scalable. We report the results of all YCSB workloads under $n=50$ in detail in Figure~\ref{fig:ycsbn50d0}. \cab consistently outperforms Raft across all workloads in terms of both throughput and latency. This advantage is particularly evident in heterogeneous settings. When compared to Raft, \cab with~\texttt{t=f10\%=5} achieves approximately $3\times$ higher throughput and $3\times$ lower latency across all workloads in heterogeneous clusters. For instance, in Workload A and F, \texttt{Cab f10\%} achieved throughputs of 27,999 and 30,864 TPS, while Raft achieved throughputs of 10,136 and 10,335 TPS, respectively. Furthermore, similar performance improvements are observed in the TPC-C workloads, as depicted in Figure~\ref{fig:tpccn50d0}. By employing a smaller failure threshold, \cab progressively enhances the overall performance for all types of transactions.

In addition, the heterogeneous cluster demonstrates superior performance compared to the homogeneous cluster under both YCSB and TPC-C, with the advantage being more pronounced in YCSB. TPC-C workloads includes certain transactions that heavily rely on locks, making it challenging to execute them in parallel. Since the heterogeneity of the deployed clusters primarily lies in the number of virtual CPUs (vCPUs) rather than hardware differences, the advantage of the heterogeneous cluster is diminished in the context of TPC-C. Nevertheless, the heterogeneity brings in a remarkable $2.3\times$ higher throughput in YCSB (Figure~\ref{fig:ycsb50d0hetero} vs.~\ref{fig:ycsb50d0homo}) and a $1.4\times$ higher throughput in TPC-C (Figure~\ref{fig:tpcc50d0hetero} vs.~\ref{fig:tpcc50d0homo}).

In the TPC-C+PostgreSQL evaluation, similar to the evaluation in YSCB, both algorithms did not encounter a notable performance drop when cluster sizes scale from $10$ to $100$. Figure~\ref{fig:tpccscales} shows the breakdown of the different types of transactions. \cab sustained the performance gain in both heterogeneous and homogeneous clusters. The performance gain obtained is proportional to the mix of the transaction types.

\textbf{Dynamic failure thresholds.}
\cab can dynamically adjust its failure threshold using its lightweight reconfiguration mechanism (\S\ref{subsubsec:reconfig}). We conducted performance evaluations using YCSB under Workload A with MongoDB, as illustrated in Figure~\ref{fig:ycsbchangingt}. Initially, $t=24$ (consistent with Raft), and subsequently decreased four times ($t=20,15,10,5$) every 20 rounds. The results indicate an increase in throughput as $t$ decreases, accompanied by a continual reduction in latency. This evaluation highlights the persistent performance gains achievable by selecting lower values for $t$.

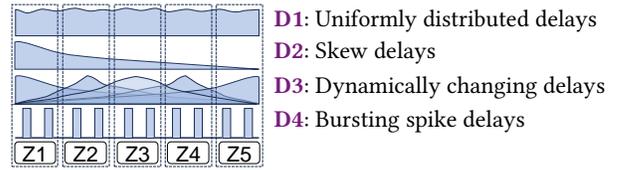
\begin{figure}
  \begin{adjustbox}{width=0.95\linewidth}
  \begin{tikzpicture}
    \node[anchor=south west, inner sep=0] (image) at (0,0) {\includegraphics[width=0.4\linewidth]{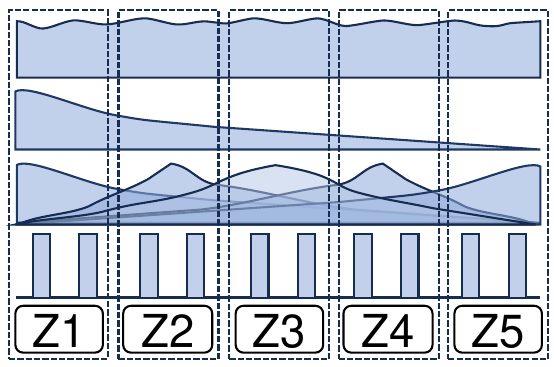}};

    \begin{scope}[x={(image.south east)},y={(image.north west)}]
      \node[anchor=north west, align=left] 
      at (1, 0.4) {\ref{d4}: Bursting spike delays};
      \node[anchor=north west, align=left] 
      at (1, 0.6) {\ref{d3}: Dynamically changing delays};
      \node[anchor=north west, align=left] 
      at (1, 0.8) {\ref{d2}: Skew delays};
      \node[anchor=north west, align=left] 
      at (1, 0.99) {\ref{d1}: Uniformly distributed delays};
    \end{scope}  
  \end{tikzpicture}
  \end{adjustbox}

  \caption{Emulated complex network conditions.}
  \label{fig:eval:delays}
  
\end{figure}

\subsection{Performance under Complex Networks}
\label{sec:eval:perfcomplexnetworks}

\begin{figure*}
\minipage{0.5\textwidth}
    \centering
    \includegraphics[width=\linewidth]{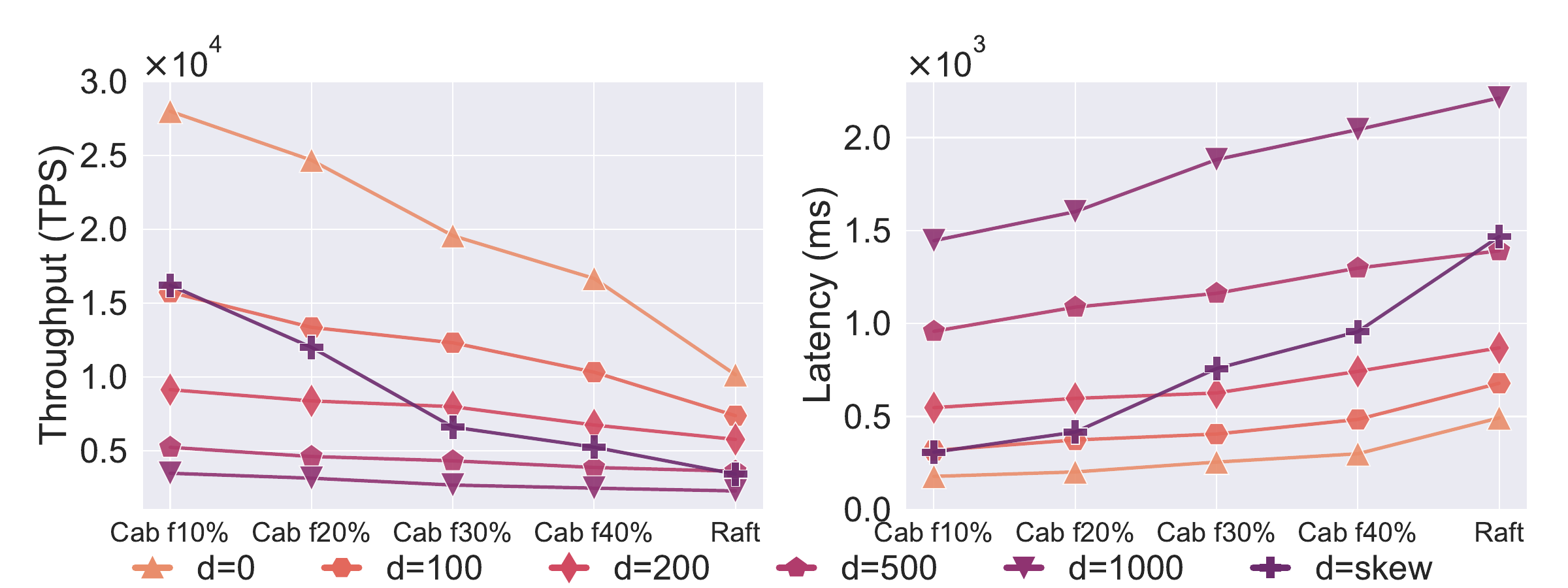}
    \subcaption{Heterogeneous cluster.}
    \label{fig:ycsbdelayshetero}
\endminipage \hfill
\minipage{0.5\textwidth}
    \centering
    \includegraphics[width=\linewidth]{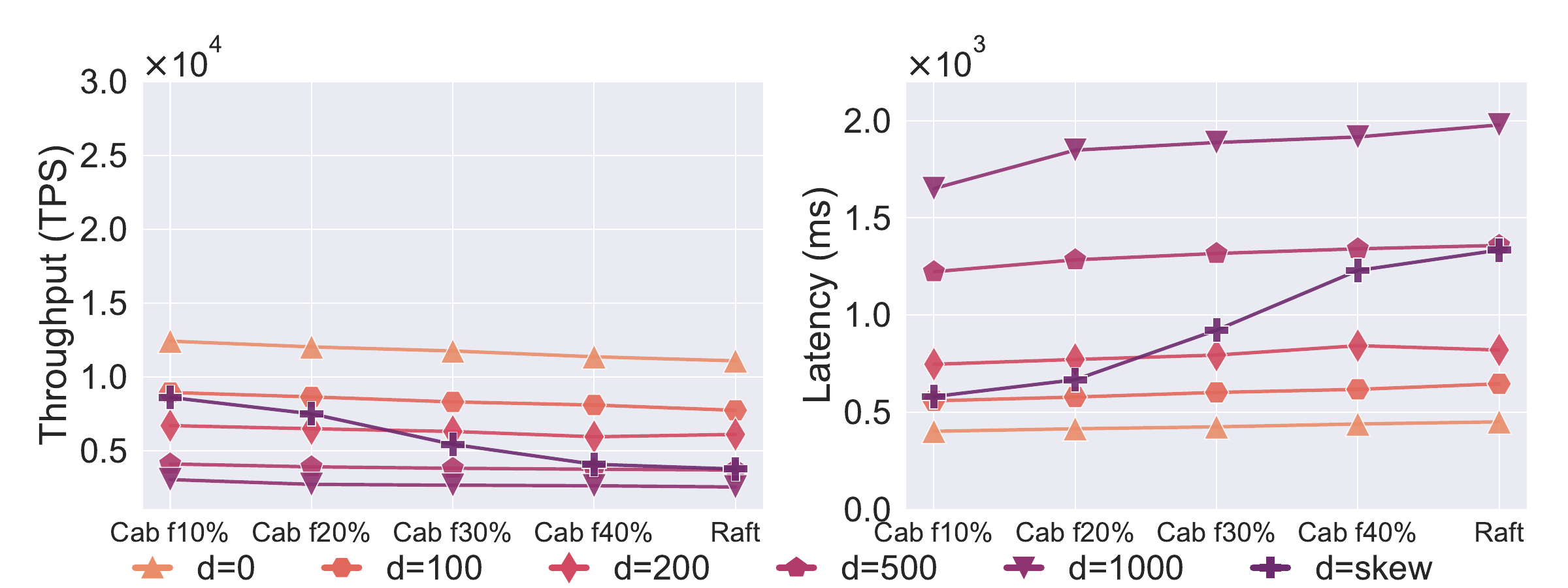}
    \subcaption{Homogeneous cluster.}
    \label{fig:ycsbdelayshomo}
\endminipage
\caption{Performance of YCSB+MongoDB under $d=\ref{d1}, \ref{d2}$ with $n=50$, $b=5k$ under Workload A.}
\label{fig:ycsbdelays}
\end{figure*}

\begin{figure*}
\minipage{0.5\textwidth}
    \centering
    \includegraphics[width=\linewidth]{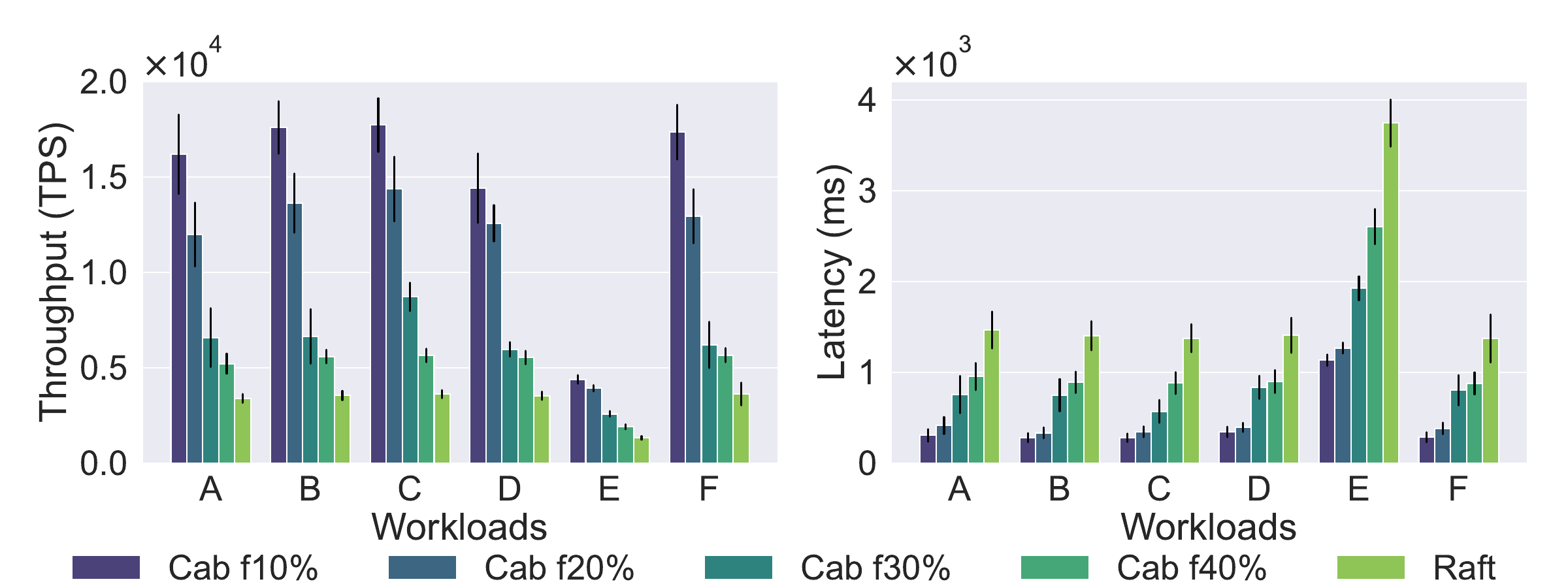}
    \subcaption{Heterogeneous cluster.}
    \label{fig:ycsbskewhetero}
\endminipage \hfill
\minipage{0.5\textwidth}
    \centering
    \includegraphics[width=\linewidth]{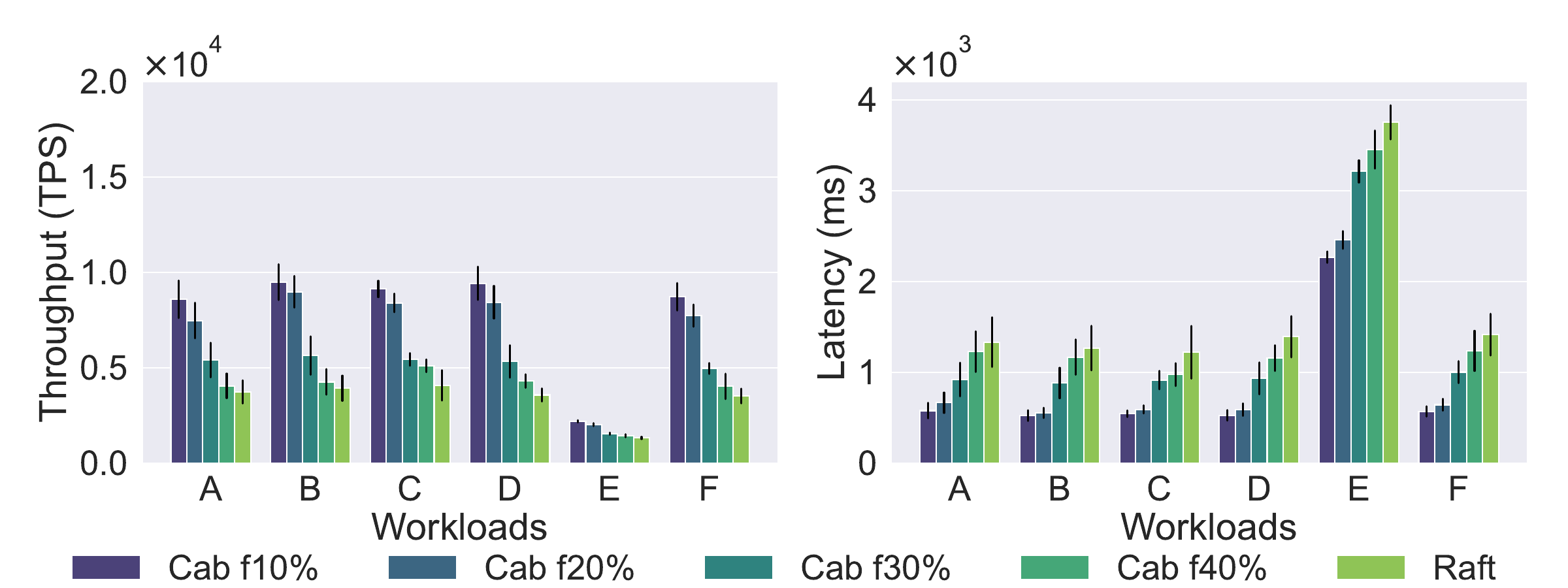}
    \subcaption{Homogeneous cluster.}
    \label{fig:ycsbskewhomo}
\endminipage
\caption{Performance of YCSB+MongoDB under skew latency ($d=\ref{d2}$) with $n=50$, $b=5k$ under all YCSB workloads.}
\label{fig:ycsbskew}
\end{figure*}

To assess the performance of \cab and Raft in complex network scenarios, we introduced additional network delays using \texttt{netem}~\cite{netem}. These delays were categorized as follows:

\begin{enumerate} [label=\textbf{D\arabic*}]  
    \item \label{d1} \textbf{Uniformly distributed delays.} The emulated delays are evenly implemented on all nodes, comprising four sets of increasing delays with variances: $d=100\pm20$~ms, $200\pm40$~ms, $500\pm100$~ms, $1000\pm200$~ms.
    
    \item \label{d2} \textbf{Skew delays.} The delays were implemented in a skewed and uneven manner across the nodes. The delay values declined from $1,000\pm200$~ms to $100\pm20$~ms progressing across the nodes (illustrated in Figure~\ref{fig:eval:delays}).
    
    \item \label{d3} \textbf{Dynamically changing delays.} The delays imposed in \ref{d2} undergo periodic changes. The intervals of these changes were carefully configured to ensure that each zone experienced the full range of emulated delay values, encompassing both the highest and lowest delays.
    
    \item \label{d4} \textbf{Bursting delays.} The delays in each zone repeatedly spiked for a brief period before dissipating. This behavior emulates heavily unstable networks.
\end{enumerate}

Compared to normal cases without network delays (as discussed in \S\ref{sec:eval:perfscalingcluster}), \cab demonstrated even greater superiority over Raft, with significantly improved throughput and latency. Figure~\ref{fig:ycsbdelays} illustrates the performance of \cab and Raft under network delays of $d=\ref{d1}$ and \ref{d2} in clusters of $n=50$. 

Under increased network delays, both \cab and Raft experienced performance degradation. Nonetheless, compared to Raft, \cab demonstrated strong resilience, with a lower increase in latency in both homogeneous and heterogeneous clusters. In addition, under $\ref{d2}$, the performance of \texttt{Cab f10\%} remained comparable to that under $\ref{d1}=100$ms, whereas Raft's performance was degraded to the level observed under $\ref{d1}=500$ms. These results indicate \cab's performance gains in diverse network conditions.

Figure~\ref{fig:ycsbskew} shows the performance of all YCSB workloads under a skew delay of~\ref{d2}, further highlighting \cab's superiority in heterogeneity. For example, \cab achieved approximately $6\times$ higher throughput (e.g., 18,899 TPS of \texttt{Cab f10\%} vs. 3045 TPS of Raft in Workload A) and $5.5\times$ lower latency (e.g., 264ms of \texttt{Cab f10\%} vs. 1477ms of Raft in Workload C). 

\begin{figure*}[t]
\minipage{0.5\textwidth}
    \centering
    \includegraphics[width=\linewidth]{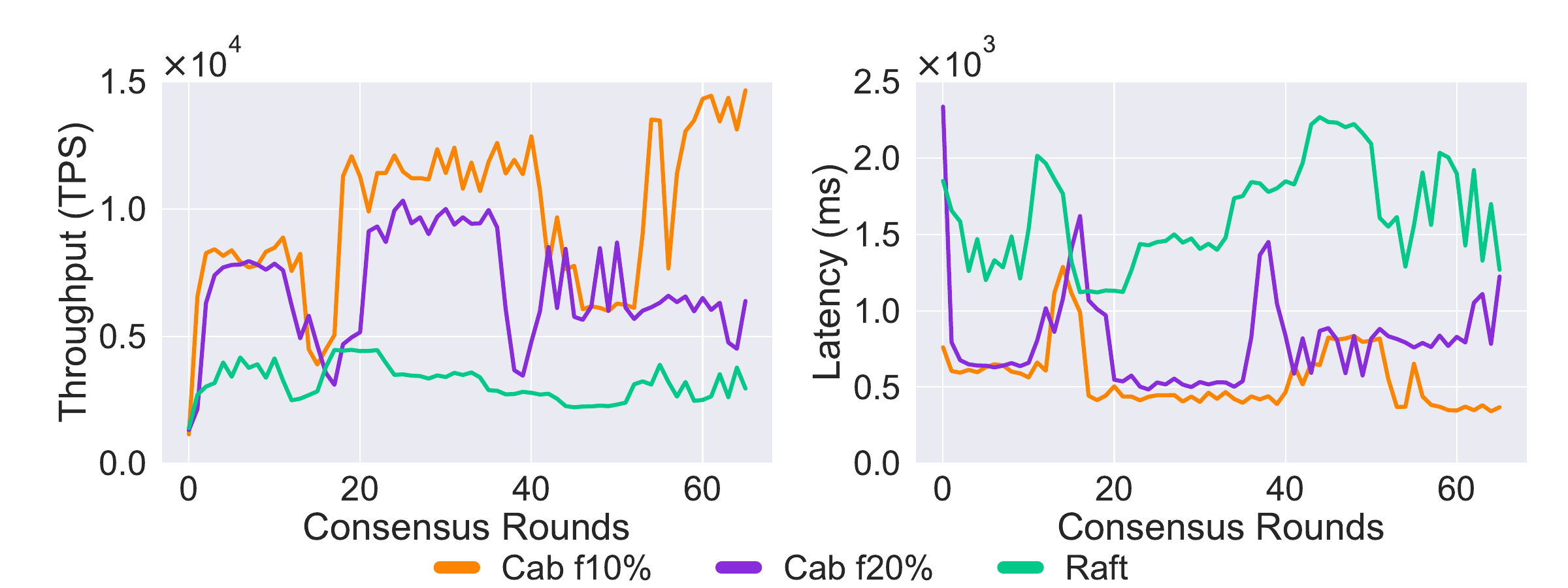}
    \subcaption{Heterogeneous cluster.}
    \label{fig:ycsbdynamicdelayshetero}
\endminipage \hfill
\minipage{0.5\textwidth}
    \centering
    \includegraphics[width=\linewidth]{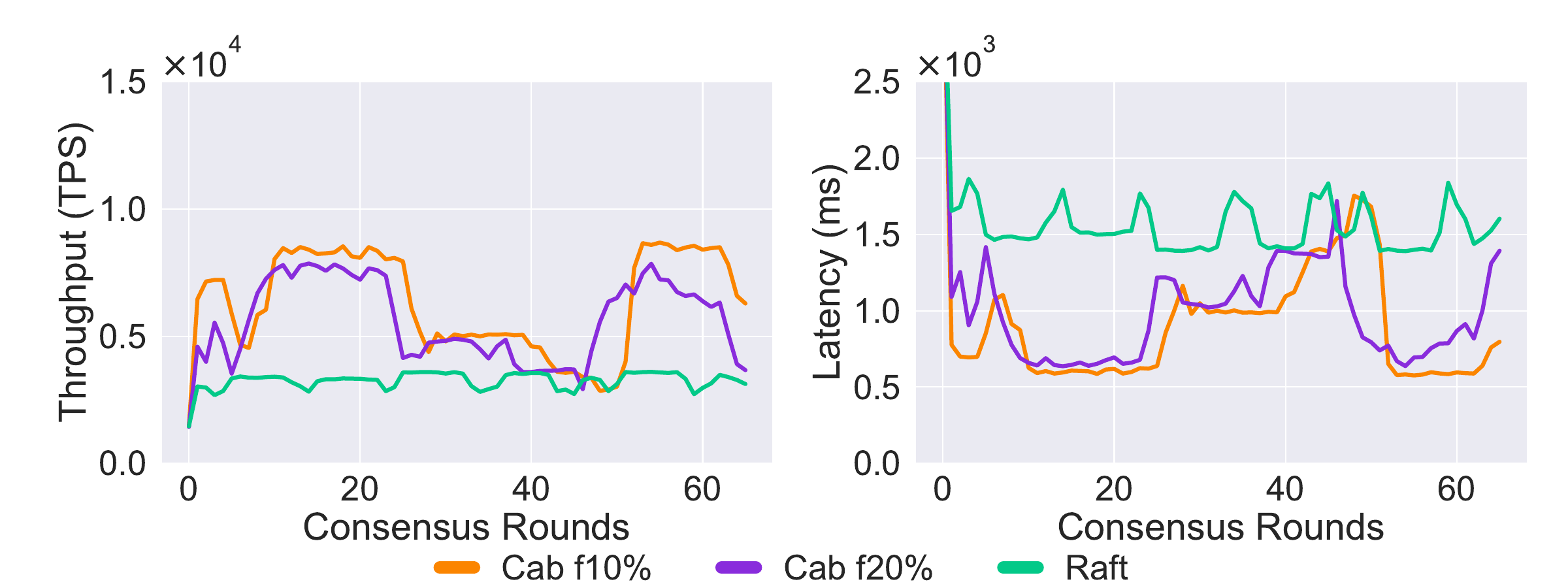}
    \subcaption{Homogeneous cluster.}
    \label{fig:ycsbdynamicdelayshomo}
\endminipage
\caption{Real-time performance of YCSB+MongoDB under $d=\ref{d3}$, $n=50$, $b=5k$ under Workload A.}
\label{fig:ycsbdynamicdelays}

\minipage{0.5\textwidth}
    \centering
    \includegraphics[width=\linewidth]{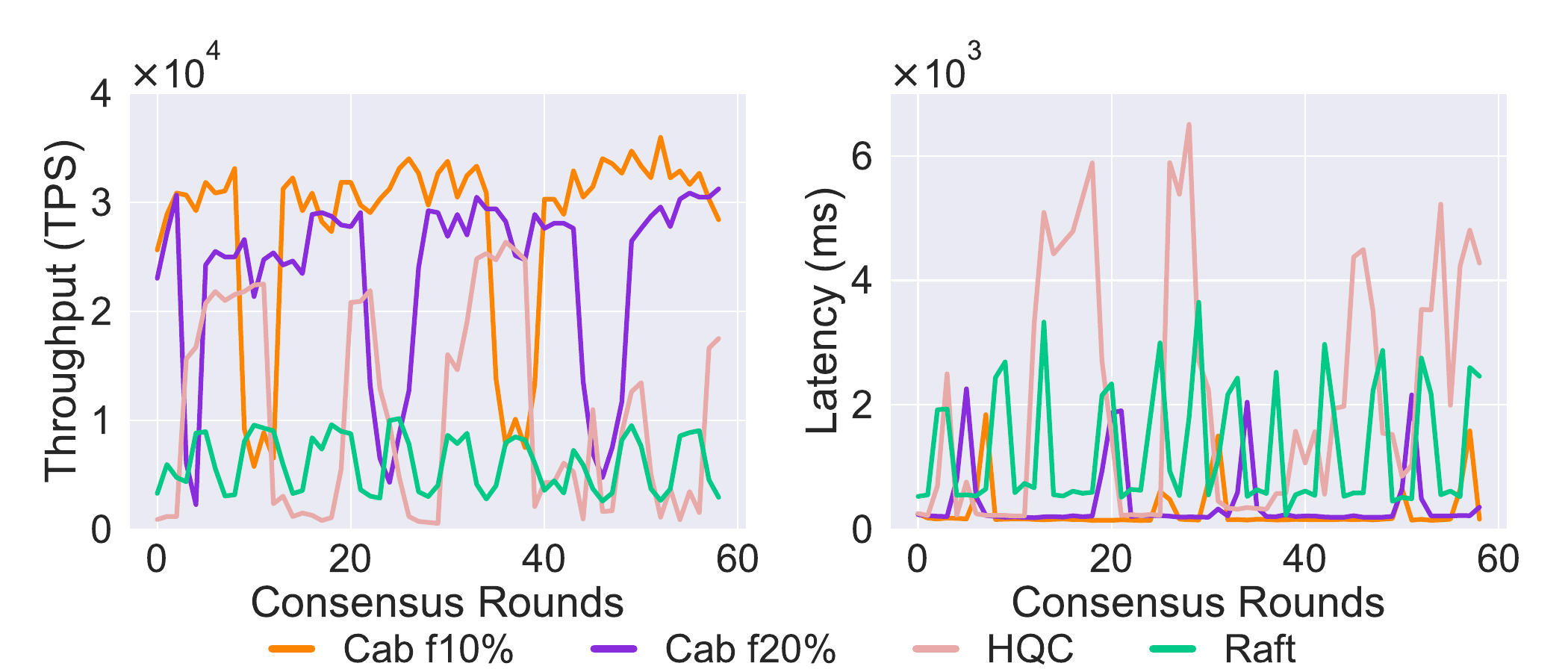}
    \subcaption{Heterogeneous cluster.}
    \label{fig:ycsbspikehetero}
\endminipage \hfill
\minipage{0.5\textwidth}
    \centering
    \includegraphics[width=\linewidth]{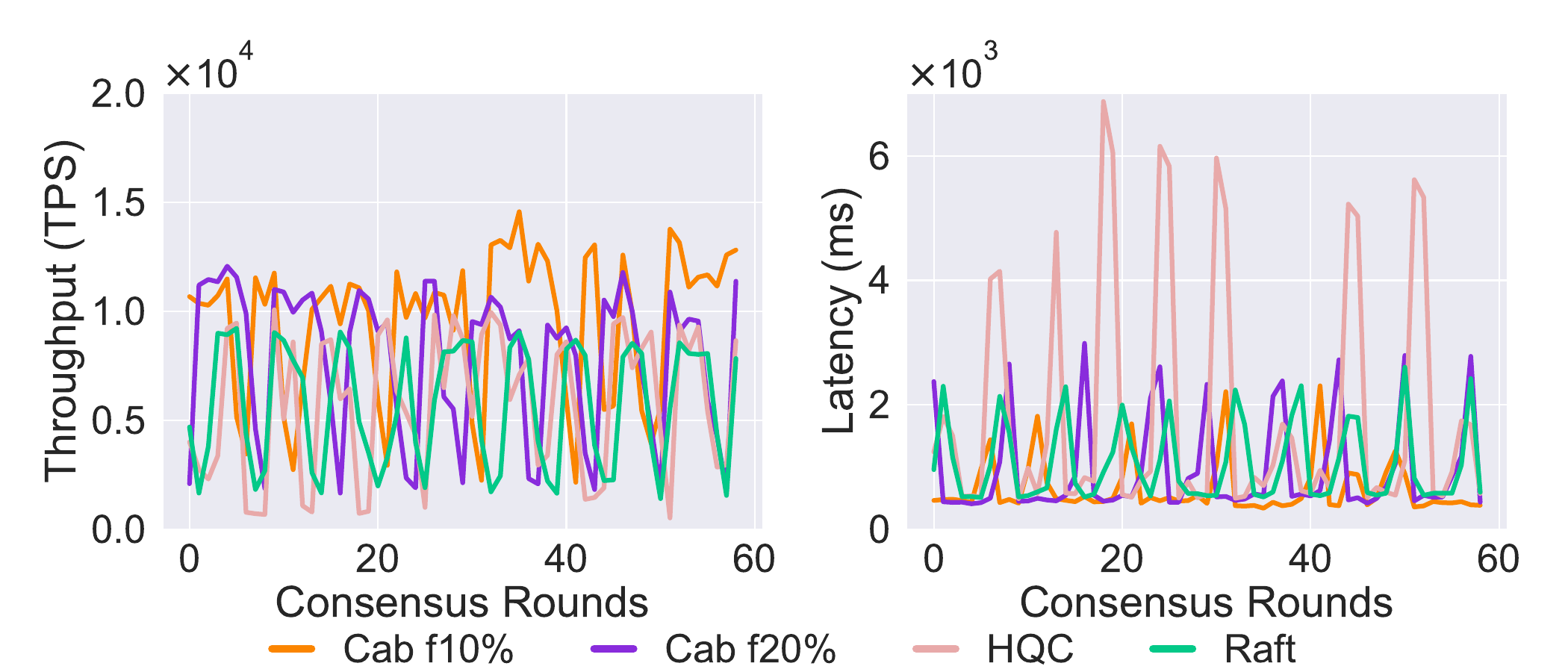}
    \subcaption{Homogeneous cluster.}
    \label{fig:ycsbspikehomo}
\endminipage

\caption{Performance of YCSB+MongoDB under $n=11$, $d=\ref{d4}$, $b=5k$ under Workload A; HQC divides cluster into 3-3-5.}
\label{fig:ycsbspike}
\end{figure*}

In addition to its high performance under static delays, \cab can effectively adapt to changing network delays, ensuring the system remains at optimal performance. Figure~\ref{fig:ycsbdynamicdelays} shows the real-time performance of \cab and Raft under varying network delays of \ref{d3}. \cab's throughput drops and latency increases (around Round 20 and 50) when strong nodes with high weights experience higher delays. Nevertheless, \cab promptly responds by reassigning high weights to currently faster nodes and regains high performance. 

\begin{figure}[t]
    \centering
    \includegraphics[width=0.99\linewidth]{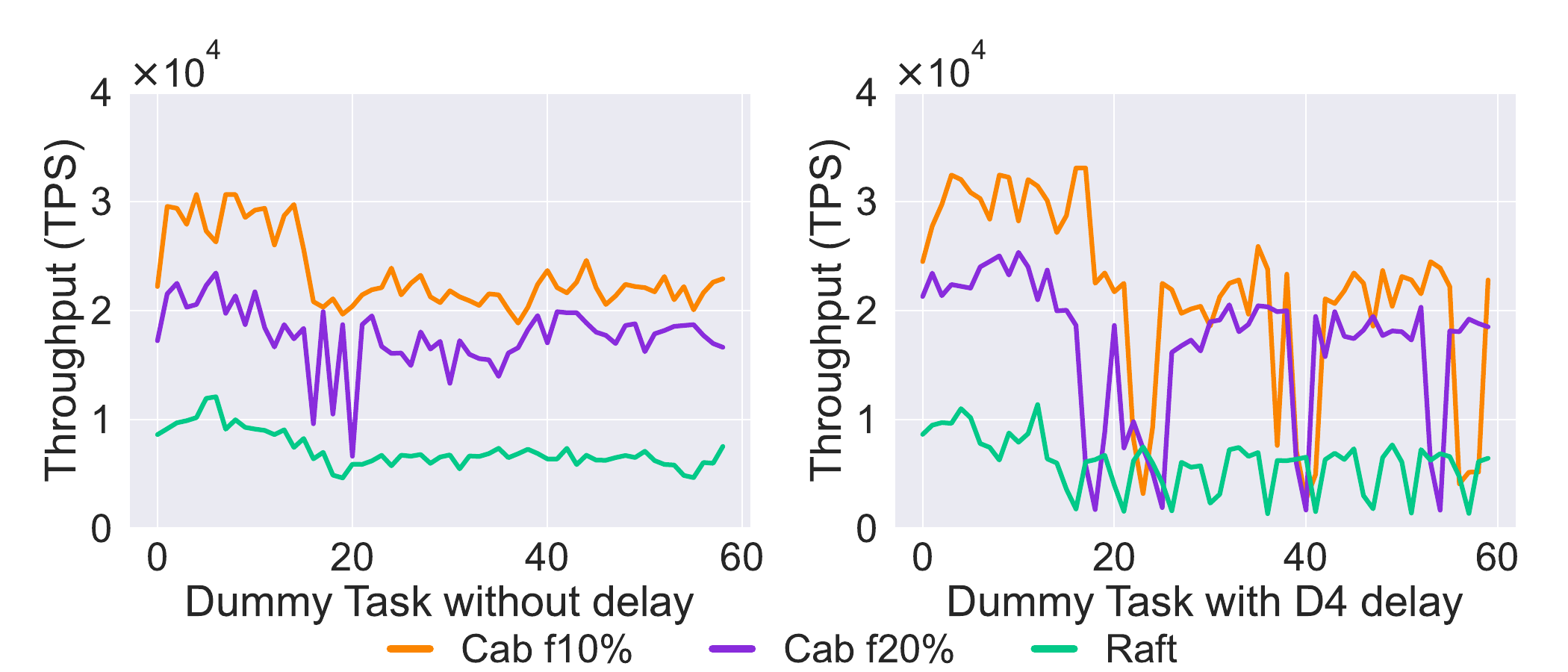}
    \caption{Real-time performance of YCSB under $n=11$, $b=5k$ under Workload A; dummy tasks start at Round 20.}
    \label{fig:dummytasks}
\end{figure}

\textbf{Bursting delays.}
In addition to dynamically changing delays, we evaluated the system's performance under bursting delay conditions (\ref{d4}), where delay spikes of 1000$\pm$100ms occur intermittently. A 5-second period of bursting delays follows a 10-second no-delay period (2:1 ratio). Figure~\ref{fig:ycsbspike} presents the real-time throughput and latency performance in both heterogeneous and homogeneous clusters. For comparison, we included hierarchical quorum consensus (HQC) as a baseline. HQC partitions the system into three groups with sizes $|G_1|=3$, $|G_2|=3$, and $|G_3|=5$. Consensus is first reached within each group, followed by global consensus among the group leaders.

\begin{figure*}[t]
    \minipage{0.499\textwidth}
    \centering
    \includegraphics[width=\linewidth]{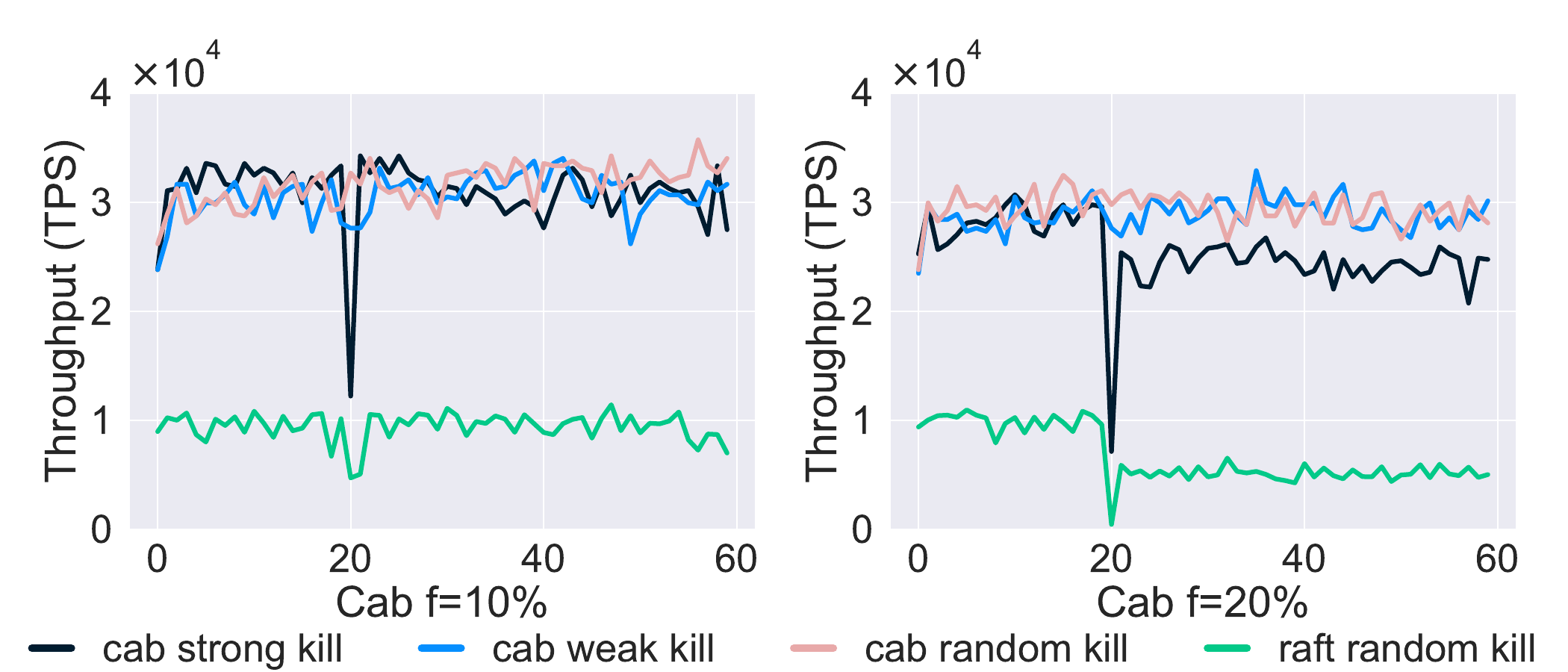}
    \subcaption{Crash failures, starting at Round 20.}
    \label{fig:crashonly}
    \endminipage \hfill
    \minipage{0.499\textwidth}
    \centering
    \includegraphics[width=\linewidth]{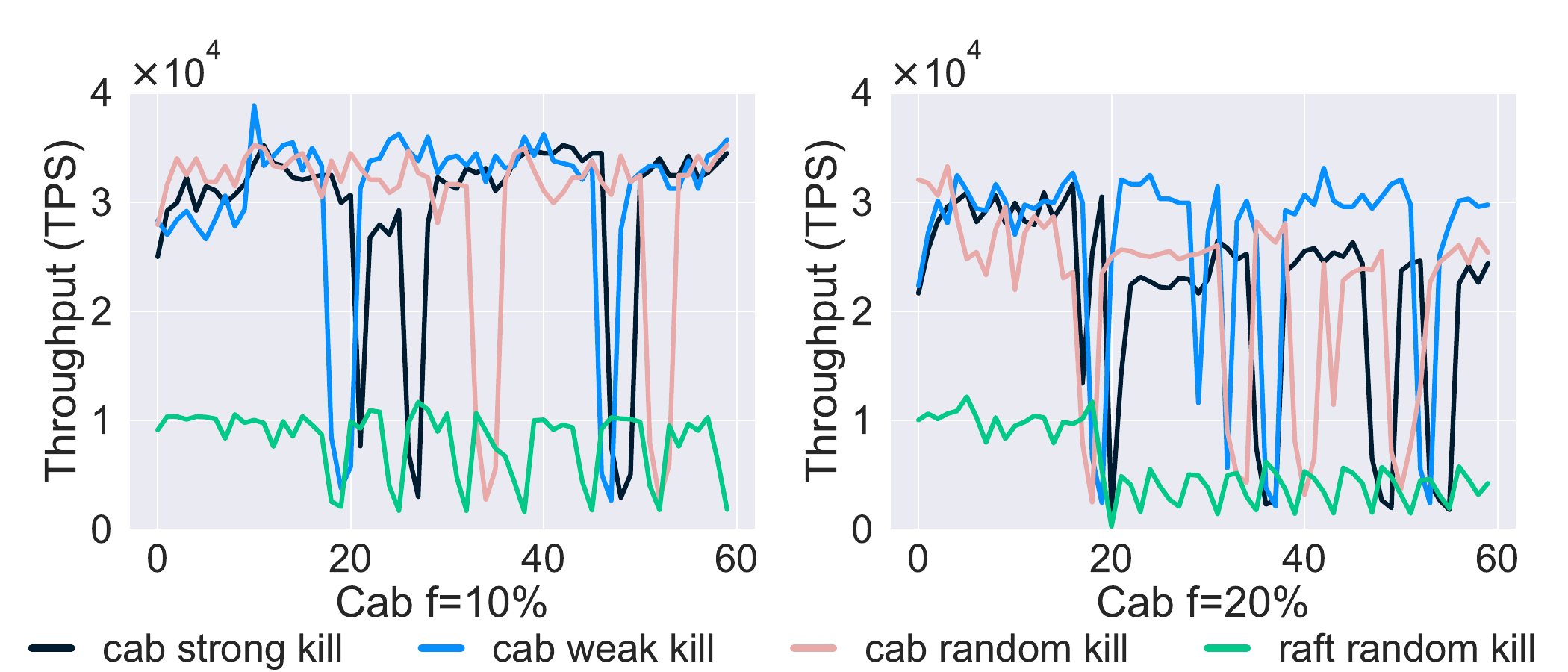}
    \subcaption{Crash failures + bursting delays (\ref{d4}), starting at Round 20.}
    \label{fig:spikeandcrash}
    \endminipage 
    \caption{Performance of crash + spike (\ref{d4}) under YCSB+MongoDB, where $n=11$, $b=5k$ under Workload A.}
    \label{fig:failures}
\end{figure*}

Under spike delays, the performance of all evaluated consensus algorithms degrades, but heterogeneous clusters demonstrate greater robustness. \cab effectively mitigates the impact of spike delays by reassigning higher weights to more responsive nodes. As analyzed in \S\ref{sec:background}, HQC incurs significantly higher latency due to its multi-round message-passing, which amplifies delays under heavy delay conditions. For example, in Figure~\ref{fig:ycsbspikehetero}, in Round 18, HQC exhibits latency (6505ms) up to $4.3\times$ that of \texttt{Cabinet f10\%} (1502ms) and $3\times$ that of Raft (2154ms) when hit by a delay spike.

\textbf{Resource contention}. We also evaluated scenarios where nodes concurrently handle other tasks, a common setup in modern deployments where physical machines run multiple containerized applications sharing CPU resources. Specifically, we introduced a CPU-heavy dummy process that repeatedly computes hash values and utilizes multiple threads equal to the number of the node’s vCPUs, creating significant CPU loads. 

Figure~\ref{fig:dummytasks} reports the performance of YCSB+MongoDB in an $n=11$ cluster, where the dummy task starts around round 20. The left figure shows results when only the dummy task is executed, while the right figure shows results when the dummy task is executed alongside bursting delays, with the first delay spike coinciding with the task's start. Without bursting delays, all three algorithms experienced a performance dip. When bursting delays were introduced, performance exhibited additional fluctuations. Notably, \cab quickly adjusted node weights and recovered from impacted rounds. The presence of the dummy task combined with spike delays did not change the performance rankings among the evaluated algorithms, highlighting \cab's robustness in handling compounded disruptions.

Although throughput inevitably drops when strong nodes experience high network delays, \cab can quickly adapt and return to high throughput by reassigning weights in the next round. This adaptability is a crucial feature of \cab, allowing it to deliver consistently high performance in real-world deployments where networks are unpredictable. 

\subsection{Performance under Failures}
\label{sec:eval:perfuderfailures}

We evaluated the performance of \cab and Raft in the presence of failures and network delays in heterogeneous clusters of $n=11$ nodes. Figure~\ref{fig:failures} shows the real-time throughput of \cab with \texttt{f10\%}, \texttt{f20\%}, and Raft under YCSB+MongoDB (Workload A) under three crash scenarios: strong kills, weak kills, and random kills. These scenarios allow us to assess how weighted consensus performs under targeted failures, where strong and weak nodes are specifically impacted, as well as random failures. Note that since Raft does not apply weights, all crashes were random kills.

\begin{description}
    \item[Strong kills] crash $x$ nodes with the top $x$  highest weight.
    \item[Weak kills] crash $x$ nodes with the bottom $x$ lowest weight.
    \item[Random kills] crash $x$ nodes randomly regardless of weights.
\end{description}

Figure~\ref{fig:crashonly} reports the evaluation results of crash failures under different strategies. All crashes occurred at Round 20. For example, in \texttt{f=20\%} under strong kills, we crashed the nodes with the top $2$ weights at Round 20. Before Round 20, all \cab operated stably at high throughput, with strong nodes assigned higher weights. At Round 20, strong kills crashed nodes with high weights, resulting in plummeted throughput in the next round (\texttt{f=10\%}). However, higher weights were promptly reassigned to non-faulty nodes, leading to a throughput recovery in the next round. The recovered throughput became lower than the throughput before the crash, as $t$ strong nodes were lost. The recovered throughput dropped more substantially (\texttt{f=10\%} vs. \texttt{f=20\%}) under a higher $t$ as more strong nodes were killed. Despite the drop in throughput after recovery, \cab still outperformed Raft.

In contrast, under weak kills, where nodes with low weights were crashed, performance was unaffected. The throughput remained unchanged compared to normal operation before Round 20, as the leader collected replies from the same set of stronger nodes. Additionally, the throughput under random kills in \cab stayed between the strong kills and weak kills scenarios since it involved killing both strong and weak nodes.

We further evaluated \cab and Raft under a combined failure scenario: crash failures and delay spikes \ref{d4}). At Round 20, crashes occurred as per the above strategies, accompanied by a sudden burst of delays across all nodes. Figure~\ref{fig:spikeandcrash} highlights \cab's resilience in this scenario. Despite the compounded failures, \cab rapidly reassigned weights to responsive nodes and recovered throughput within a few rounds. In contrast, Raft exhibited slower recovery and lower post-failure throughput.

To conclude, \cab demonstrated strong robustness and adaptability under diverse network delays and failure conditions. These results highlight that \cab's efficiency makes it particularly suitable for large-scale, heterogeneous systems.

\section{Conclusions}
Consensus algorithms relying on majority quorums encounter inefficiencies in large-scale, especially heterogeneous, systems, as strong nodes are forced to wait for weaker nodes to reach consensus. This paper proposes \cab, which enables dynamic weighted consensus. Consensus is no longer based on a majority of physical nodes but on a majority of weights, allowing stronger nodes to have a more significant influence in the agreement process.
Furthermore, \cab dynamically reassigns high weights to more responsive nodes, ensuring the system stays at optimal performance. In our evaluation, \cab consistently outperforms Raft under various scenarios, including increasing system scales, static, skew, and dynamic network delays, and different proportions of crash failures.

\bibliographystyle{ACM-Reference-Format}
\bibliography{ref.bib}

\end{document}

%% file: AlgoAugmentedRPC.tex
{
\begin{algorithm}[t]
\algrenewcommand\algorithmicwhile{\textbf{upon}}

\algblockdefx[Loop]{MyFor}{EndMyFor}[1]{\textbf{for} #1 \textbf{do}}{\textbf{end for}}
\algnotext{EndMyFor}

\caption{\textsc{Cabinet-Consensus}}
\label{algo:cabinet}
\begin{algorithmic}[1]
  \State \textbf{struct} AppendEntriesRPC: \label{algo:RPCstruct}
    \Statex \quad $\ldots$ \quad \textit{// original parameters in Raft's AppendEntries RPCs}
    \State \quad \textbf{int} $wclock$ \quad \textit{// added by Cabinet} \label{algo:RPCwclock}
    \State \quad \textbf{float} $weight$ \quad \textit{// added by Cabinet} \label{algo:RPCweight}

  \Statex \textcolor{blue}{$\triangledown$ As a leader}
   \State $wclock :=0 $
   \State $ws := $\Call{InitWS}{$wclock$, $n$, $t$} \label{al:initws} \Comment{ Generate $ws[0] = \{ w_1, w_2, ..., w_n \}$} 
   
   \For{\textbf{each} $n_i$} \Comment{Initialize node weights according to $ws$} 
        \State \Call{UpdateWgt}{$wclock+1$, $n_i$, $ws$} \label{algo:initweights} \Comment{$n_i$'s weight is $w_i$ in $ws$}
   \EndFor
  \Procedure{WeightedConsensus}{}
    \State $wclock$++ \Comment{Increment weight clock}
    \State $wQ := []$ \Comment{A queue for receiving replies}
    \For{\textbf{each} follower $n_i$ }

        \State $w_i := \Call{GetWeight}{wclock, n_i}$ \Comment{Get $n_i$'s weight}
        \State \textbf{issue} \Call{AppendEntriesRPC}{$wclock, w_i, \&wQ, \ldots$} \label{algo:callrpc}
    \EndFor

    \State $sum := w_{\lambda}$ \Comment{$w_{\lambda}$ is the leader's weight}
    \For{$(w_i, n_i)\leftarrow wQ$} \label{algo:wq}
        \State $sum += w_i$
        \Statex {\hspace{3em} // \textit{Faster nodes are assigned with higher weights.}}

        \State \Call{UpdateWgt}{$wclock+1$, $n_i$, $ws$} 
        \label{algo:updwgtcab}

        \If{$sum > CT$} \Comment{Consensus is reached} \label{algo:consreached}
            \State \textbf{break}
        \EndIf
    \EndFor
    
    \Statex {\hspace{1.5em} // \textit{Assign weights to remaining nodes not in $wQ$.}}

    \For{\textbf{each} remaining $n_i$} \label{algo:remaining}
        \State \Call{UpdateWgt}{$wclock+1$, $n_i$, $ws$} 
    \EndFor
  \EndProcedure

\Statex {//\textit{ \cab adds two new parameters, $w_i$ and $wclock$ to the leader's \textsc{AppendEntriesRPCs}. Replies are handled in a FIFO queue, $wQ$.} 
}  
\Function{AppendEntriesRPCs}{$wclock, w_i, *wQ, \ldots$} \label{algo:augstart}
    \State RPC.\Call{newWeight}{$wclock$, $w_i$} \Comment{See Line~\ref{algo:newwgt}}
    \State RPC.\Call{raftTask}{$\dots$} \Comment{Original Raft operations}
    \State $wQ \leftarrow (w_i, n_i)$ \Comment{First reply, first enqueue} \label{algo:augend}
\EndFunction


\Statex {\small // \textit{The leader assigns weights to followers in each wclock First Come First Serve based on the executing order, where $ws[0]$ and $ws[n-1]$ are the highest and the lowest weight, respectively. In each round, initially, $k=0$.} 
}
\Function{UpdateWgt}{$wclock, n_i, ws$} \label{algo:udpwgt}
    \State $n_i = ws[wclock][k]$ 
    \State $k++$
\EndFunction

\Statex \textcolor{blue}{$\triangledown$ As a follower}
\Statex {// \textit{Store the weight clock and the weight value issued by the leader.}}
\Function{NewWeight}{$wclock, w_i$} \label{algo:newwgt}
    \State $state.wc = wclock$
    \State $state.w = w_i$
\EndFunction

\end{algorithmic}
\end{algorithm}
}